\documentstyle[aps,pra]{revtex}

\def\CC{{\rm\kern.24em \vrule width.04em height1.46ex depth-.07ex
\kern-.30em C}}

\begin{document}

\title{Suppression of decoherence in quantum registers by entanglement with a nonequilibrium environment}
\author{S. Gheorghiu-Svirschevski\footnotemark[1]\footnotetext{e-mail: hnmg@soa.com}}
\address{1087 Beacon St., Suite 301, Newton, MA 02459}
\date{\today}
\maketitle

\begin{abstract}
It is shown that a nonequilibrium environment can be instrumental in suppressing decoherence between distinct decoherence free subspaces in quantum registers. The effect is found in the framework of exact coherent-product solutions for model registers decohering in a bath of degenerate harmonic modes, through couplings linear in bath coordinates. These solutions represent a natural nonequilibrium extension of the standard solution for a decoupled initial register state and a thermal environment. Under appropriate conditions, the corresponding reduced register distribution can propagate in an unperturbed manner, even in the presence of entanglement between states belonging to distinct decoherence free subspaces, and despite persistent bath entanglement. As a byproduct, we also obtain a refined picture of coherence dynamics under bang-bang decoherence control. In particular, it is shown that each radio-frequency pulse in a typical bang-bang cycle induces a revival of coherence, and that these revivals are exploited in a natural way by the time-symmetrized version of the bang-bang protocol.     
\end{abstract}
\pacs{03.65.-w; 03.67.-a; 05.30.-d}

\section{Introduction}
\label{Sec1}
 
At a very fundamental level the control of decoherence in a quantum register amounts to the engineering of its entanglement with the surrounding environment. The earliest approach to the problem was guided by the philosophy of classical error correction codes, and has culminated into a comprehensive theory of {\it quantum error correction codes} [QECCs] \cite{QECC1,QECC2,QECC3,QECC4,QECC5,QECC6,QECC7}. In essence QECCs are designed to actively purge unwanted environment entanglement by a recurring error correction cycle, to the effect of recovering periodically an 'error-free' reduced register state. Other {\it active }strategies, known as {\it quantum bang-bang control } or {\it dynamical decoupling} \cite{QBB,dyn-dcpl1,dyn-dcpl2,dyn-dcpl3,Viola,dyn-dcpl4,dyn-dcpl5}, have been developed by analogy with NMR decoupling and refocusing techniques. In this case entanglement with the environment need not necessarily cancel at the end of the correction cycle, but is coerced, by suitably engineered control "pulses", to "average out" in such a way as to leave the reduced register state virtually unchanged \cite{note}. Alternatively, it was shown that specific forms of decoherence can be {\it passively} suppressed, provided the intrinsic symmetries of the noise-generating interactions allow the existence in the register Hilbert space of {\it decoherence-free subspaces} [DFSs] that are inaccesible to noise processes \cite{DFS-Zanardi,DFS-Duan,DFS-Lidar1,DFS-Lidar2,DFS-Lidar3}. Since any entanglement with the environment leaves a DFS code unaffected, robust storage of the reduced register state can be achieved without further active correction cycles. The concept of DFS has emerged recently as only a particular instance of symmetry generated {\it noiseless subsystem }[NS] structures, capable to support robust passive storage through redundant encoding \cite{NS}. Both types of structures have been confirmed experimentally, through the implementation of optical DF states \cite{DFS-exp/opt}, of one-qubit DFS memories in two trapped ions \cite{DFS-exp/ions}, or in two nuclear spins \cite{DFS-exp/NMR}, and of a one-qubit NS memory encoded in three nuclear spins \cite{NS-exp}. Notably, QECCs and dynamical decoupling strategies have been also shown to enforce symmetrized effective evolutions capable to support NSs, so that noise-protected encoding takes place in fact in dynamically generated NSs \cite{NS-dyn}. From a broader perspective, in all cases decoherence control aims to maintain, or recover periodically, a given reduced register state in the presence of time-dependent entanglement with environmental modes. The ability to implement such effective noise protection strategies is paramount for the practical development of both robust quantum memories and fault-tolerant quantum computation.   

For a typical model register of N two-level elements, immersed in a quantized environment and described by 

\begin{equation}
\label{eq1}
H = \sum\limits_{n = 0}^{N - 1} {\varepsilon \sigma _z^{(n)} }  + \sum\limits_q {\hbar \omega _q b_q^ {\dagger}  b_q }  + \sum\limits_{n = 0}^{N - 1} {\sum\limits_q {\sigma _z^{(n)} \left( {\chi _{qn} b_q^{\dagger}   + \chi _{qn}^*b_q}\right)}} \;\;,
\end{equation}

\noindent the passive or active character of the control strategy seems to be conditioned by the number of DFSs accessed by the encoding method. Indeed, since the associated Hilbert space is always decomposable into a direct sum of elementary DFSs [at the very least, direct products of the 1-dimensional eigenspaces for each $\sigma _z^{(n)} $], the corresponding reduced register state is necessarily distributed over a [finite] number of disjoint DFSs. If the encoding state is confined to a single DFS, decoherence is passively suppressed, even under entanglement with the environment. But if the encoding involves multiple entangled DFSs, active control seems mandatory. 

The main intention of this paper is to point out that decoherence-free states, and therefore passive control, may be possible as well for nonseparable distributions entangled over multiple DFS, and also entangled with the environment. Such distributions realize a noise-protected concatenation of distinct DFS, which suggests, in principle, that the capacity for coherent storing of a DFS-supporting register may extend beyond the dimensionality of the largest DFS. Sec.II presents a general argument in favor of this possibility. For a more precise discussion, we exploit an extended class of exact, closed-form density matrix solutions for systems with decoherence-free subspaces under interactions linear in environmental degrees of freedom, which have been retrieved as density matrix generalizations of the pure state coherent-product [Davydov] ansatz in ref.\cite{mine}. After providing some straightforward Davydov [pure] state examples of decoherence-free propagation on multiple DFS, in Sec.III we introduce the thermal coherent-product states in a self-contained manner, and with reference to the specific context of quantum registers. These states involve a nonequilibrium environment in a statistical superposition of Gaussian states and provide a natural extension of the standard solution currently employed in discussions of the decoherence process \cite{decoh,decoh1}, which refers to an uncorrelated thermal environment. We show in Sec.IV that the corresponding register reduced  density matrix can propagate, under proper conditions, in an unperturbed, decoherence free manner, even when nontrivially distributed over multiple, nonisomorphic DFS, and strongly entangled with the environment. In Sec.V we discuss the conditions for multi-DFS decoherence free evolution for three particular cases of the spin-boson model (\ref{eq1}): a single-qubit register, a multi-qubit register with weak collective decoherence and a multi-qubit register with individual decoherence. In Sec.VI it is pointed out that sudden spin-flips of the type considered in bang-bang protocols \cite{QBB} imply in effect a manipulation of the register-environment entanglement that may be of interest for the preparation of multi-DFS decoherence free states. Alternatively, it is also found that the density matrix coherent-product solutions for model (\ref{eq1}) prove instrumental for an analysis of the detailed coherence dynamics under bang-bang control \cite{QBB}. It is shown that each qubit spin-flip in a bang-bang cycle produces a revival of coherence, and that these revivals provide an alternative justification for the well-known superior performance of the time-symmetric version of the protocol \cite{Viola}. A summary and conclusion are provided in Sec.VII.

\section{The general model}

Consider a quantum register R, described by a Hamiltonian $H_R$, and immersed in an environment B with Hamiltonian $H_B$. Let the register-environment interaction be given as $H_{SB} = \sum\limits_k {R_k \otimes B_k}$, with $R_k$ [$B_k$] self-adjoint register [environment] coupling operators, such that the total Hamiltonian reads

\begin{equation}
\label{eq2}
H = H_R  + H_B  + \sum\limits_k {R_k \otimes B_k}\;. 
\end{equation}   

\noindent Let us assume also that the register dynamics supports multiple decoherence-free subspaces ${\cal S}_J$ labeled by an appropriate index set $\{J\}$. Here we refer to the closed-system, hamiltonian version of the theory of DFSs \cite{DFS-Zanardi,DFS-Lidar2}. That is, let ${\cal A}$ denote the $\dagger$-closed register interaction algebra generated by the hamiltonian $H_R$, the register components of the interaction $R_k$, and the identity $I_R$ on the register Hilbert space. By a basic theorem of $\dagger$-closed operator algebras, ${\cal A}$ decomposes into a direct sum of $d_J \times d_J$-dimensional irreps, each of multiplicity $n_J$, i.e.

\begin{mathletters}
\label{eq2a}

\begin{equation}
{\cal A} = \mathop \bigoplus \limits_J I_{n_J} \otimes {\cal M} (d_J \times d_J,\CC)\;,
\end{equation}

\noindent where ${\cal M} (d_J \times d_J,\CC)$ denotes the algebra of $d_J \times d_J$ complex matrices. As a result, the register Hilbert space ${\cal H}_R$ also decomposes into a direct sum of subspaces invariant under ${\cal A}$, as

\begin{equation}
{\cal H}_R = \bigoplus \limits_J \CC^{n_J}\otimes\CC^{d_J}\;\;.
\end{equation}

\end{mathletters}

\noindent The register DFSs ${\cal S}_J$ correspond then to the 1-dimensional irreps ${\cal M} (1 \times 1,\CC)$ of ${\cal A}$, and their dimension is identical to the multiplicity of the associated irrep, ${\rm dim}{\cal S}_J = n_J$. Equivalently, a necessary and sufficient condition for the occurrence of DFSs \cite{mine} is that the kernels of all commutators $[H_R, R_k]$ have a nontrivial intersection [e.g., $[H_R, R_k]=0$, as in model (\ref{eq1})]. Further, 

i) any register state $|\psi^R_J\rangle$ belonging to the $J$-th DFS,  $|\psi^R_J\rangle \in {\cal S}_J$, is a simultaneous eigenstate of all register couplings $R_k$ for real eigenvalues $\mu_{k,J}$ common to the entire DFS, 

\[
R_k |\psi^R_J\rangle = \mu_{k,J} |\psi^R_J\rangle\;\; ;
\]
 
ii) the DFS is left invariant by the register hamiltonian $H_R$, 

\[
H_R|\psi^R_J\rangle \in {\cal S}_J\;\;.
\]

\noindent A ubiquitous example of register supporting DFSs is provided by a spin-boson model (\ref{eq1}) with {\it weak collective decoherence}, i.e., with identical coupling of each spin to a common bath, via [e.g.] the $\sigma_z^{(n)}$ component only, such that $\chi_{qn} \equiv \chi_q$ [the more general {\it collective decoherence }case involves all spin components, through interactions of the form $\vec{\sigma}^{(n)}\cdot \vec{\chi}_q$].  In this case the register-bath interaction reduces to  $S_z {\sum\limits_q \left( {\chi _{q} b_q^{\dagger}   + \chi _{q}^*b_q}\right)}$, where $S_z = \sum\limits_{n = 0}^{N - 1} {\sigma _z^{(n)}}$ denotes the z-component of the total spin, and the DFSs are the [highly] degenerate eigenspaces of $S_z$. 

Returning now to our general model, throughout the following we shall restrict the register state space to 

\[
{\cal R} =\mathop \bigoplus \limits_J {\cal S}_J\;\;,
\]

\noindent such that if ${\cal B}$ is the state space of the environment, the joint register-environment states live in the  direct product Hilbert space ${\cal H} = {\cal R} \otimes {\cal B}$. In this case, the subalgebra of ${\cal A}$ generated by the restrictions of $H_R$, $R_k$ and $I_R$ on ${\cal R}$ is necessarily abelian. Hence, if $P_J$ are the self-adjoint, orthogonal projectors on the DF subspaces ${\cal S}_J$, such that $P_J P_K = P_K  P_J = \delta_{J,K}$, and $\sum\limits_J {P_J} = I_{\cal R}$,  where $I_{\cal R}$ is now the identity on ${\cal R}$, properties (i-ii) above show that 

i') every $R_k$ has on $\cal R$ the simple form

\begin{equation}
\label{eq3}
R_k = \sum\limits_J {\mu_{k,J}P_J}\;\;,
\end{equation}

ii') $[H_R, P_J] = 0$ and, accordingly, the total Hamiltonian (\ref{eq2}) describes a quasi-separable dynamics on ${\cal R}$. That is, it generates a superposition of independent evolutions on individual DFSs, where each DFS creates its own separable environment. Indeed, Hamiltonian (\ref{eq2}) can be written, by making use of the orthogonality and completeness of the projectors $\{ P_J \}$ on ${\cal R}$, 

\[
H = \sum\limits_J {  \left[{ P_J H_R P_J + P_J \otimes \left( H_B +  \sum\limits_k{\mu_{k, J} B_k }\right)  }\right] }\;\;,
\]

\noindent and the evolution it generates obviously factores into separate evolutions on the individual DFS. 

Our starting observation is that, alternatively, the entire unperturbed register evolution can be formally factored out of the total evolution. It is sufficient to note that the register Hamiltonian $H_R = \sum\limits_J { P_J H_J P_J}$ commutes with each of the environment driven terms in the sum. Then the evolution operator generated by $H$ can be conveniently cast as

\[
\exp\left[{ -\frac{i}{\hbar}Ht }\right] = \exp\left[{ -\frac{i}{\hbar}H_Rt }\right] \cdot \exp\left[{ -\frac{i}{\hbar} \sum\limits_J {P_J \otimes H_{B,J} } }\right] = 
\]

\begin{equation}
\label{eq4}
 = \exp\left[{ -\frac{i}{\hbar}H_Rt }\right] \cdot \sum\limits_J \exp\left[{ -\frac{i}{\hbar}H_{B,J}t }\right]  \otimes P_J\;\;,
\end{equation} 

\noindent with

\begin{equation}
\label{eq3b}
H_{B,J} = H_B + \sum\limits_k{\mu_{k, J} B_k }\;\;.
\end{equation}

\noindent This factorization hints, in a rather evident manner, to the possibility of unperturbed and decoherence free register dynamics on multiple DFS, since the register reduced density matrix at time $t$ acquires the form

\begin{mathletters}
\label{eq5}

\begin{equation}
\hat\rho^R(t) = \exp\left[{ -\frac{i}{\hbar}H_Rt }\right] \cdot \hat{\overline\rho}^R(t) \cdot \exp\left[{\frac{i}{\hbar}H_Rt }\right] \;\;,
\end{equation}

\noindent with

\begin{equation}
\hat{\overline\rho}^R(t) = Tr_B \left[ { \sum\limits_{J, J'} \exp\left[{ -\frac{i}{\hbar}H_{B,J}t }\right] \cdot \hat\rho(0) \cdot  \exp\left[{ \frac{i}{\hbar}H_{B,J'}t }\right] }\right] \;\;.
\end{equation}

\end{mathletters}

\noindent When the initial total density matrix $\hat\rho(0)$ is such that at all times
\[
\hat{\overline\rho}^R(t) =\hat\rho^R(0)\;\;,
\]

\noindent the register evolution becomes indeed unperturbed and decoherence free, 

\[
\hat \rho^R(t) =  \exp\left[{ -\frac{i}{\hbar}H_Rt }\right] \cdot \hat\rho^R(0) \cdot \exp\left[{ \frac{i}{\hbar}H_Rt }\right] \;\;.
\]

\noindent However, there is no indication yet that the conditions for this peculiar dynamics can be met outside of individual DFS. For a better picture, let the initial density matrix $\hat\rho(0)$ be given a representation in terms of a tensor product basis in $\cal H$, of the form $\{|\phi^R_{J\nu}\rangle \otimes |\phi^B_{\sigma}\rangle\}$, where $\{|\phi^B_{\sigma}\rangle\}$ is an orthonormal basis in $\cal B$ and $\{|\phi^R_{J\nu}\rangle\}$ an orthonormal register basis yielding an irreducible representation [irrep] on $\cal R$, with  $\nu$ an index labeling basis states belonging to the same DFS. That is, let 

\[
\hat \rho(0) = \sum\limits_{J\nu,\sigma}{\sum\limits_{J'\nu',\sigma'}{|\phi^R_{J\nu}\rangle \otimes |\phi^B_{\sigma}\rangle \rho_{J\nu,\sigma; J'\nu',\sigma' }(0) \langle\phi^B_{\sigma'}| \otimes \langle\phi^R_{J'\nu'}|}}\;\;.
\]

\noindent Then the corresponding reduced density matrix for the register [Eq.(\ref{eq5}b)] can be brought to the more transparent form

\begin{mathletters}
\label{eq6}

\begin{equation}
\hat \rho^R(t) =  \exp\left[{ -\frac{i}{\hbar}H_Rt }\right] \cdot \left[{ \sum\limits_{J\nu}{\sum\limits_{J'\nu'}{|\phi^R_{J\nu}\rangle \lambda_{J\nu,J'\nu'}(t) \langle\phi^R_{J'\nu'}| }} }\right]\cdot \exp\left[{ \frac{i}{\hbar}H_Rt }\right] \;\;,
\end{equation}

\noindent where the environment driven correlation amplitudes $\lambda_{J\nu,J'\nu'}(t)$ are given by

\begin{equation}
\lambda_{J\nu,J'\nu'}(t) = \sum\limits_{\sigma,\;\sigma'} { \rho_{J\nu,\sigma; J'\nu',\sigma' }(0) \langle\phi^B_{\sigma'}| \exp\left[{\frac{i}{\hbar}H_{B,J} t }\right] \cdot \exp\left[{ -\frac{i}{\hbar}H_{B,J'} t }\right] |\phi^B_{\sigma}\rangle }  \;\;.
\end{equation}

\end{mathletters}

\noindent Expression (\ref{eq6}a) now shows in fair detail that the reduced register evolution can become unperturbed, {\it even in the presence of environment-mediated entanglement between distinct DFS [$J\neq J'$]}, provided the correlation amplitudes $\lambda_{J\nu,J'\nu'}(t)$ remain constant in time. Note that any correlations $\lambda_{J\nu,J\nu'}$ between register states in the same DFS are always constant in time, and the projected dynamics $P_J \hat\rho_R(t) P_J$ on any DFS is efectively decoupled from the environment. This is a direct expression of the quasi-separable nature of Hamiltonian (\ref{eq2}). 

It is also worth pointing out that, by construction, such multi-DFS decoherence free states are distinct from distributions on noiseless subsystems \cite{NS}. The latter are defined by nonabelian, multidimensional irreps ${\cal M} (d_j \times d_J, \CC)$ of the register interaction algebra ${\cal A}$ [see Eq.(\ref{eq2a}a)] and, according to decomposition (\ref{eq2a}b), reside in the subspaces $\CC^{n_J} \otimes \CC^{d_J}$, $d_J > 1$, of the total register Hilbert space. Note that a DFS is in effect a trivial noiseless subsystem, supported by a direct sum of equivalent [identical] 1-dimensional irreps of the interaction algebra. In contrast, the multi-DFS decoherence free states referred to above are defined on a direct sum of presumably non-isomorphic DFS, i.e., on a direct sum of non-identical 1-dimensional irreps of the register interaction algebra. The register [sub]space ${\cal R}$ carrying these states is in fact orthogonal to any subspace corresponding to a nontrivial noiseless subsystem.\\

\section{The coherent product states}
\subsection{The model environment}

From the expression of the density matrix factor (\ref{eq5}b), or equivalently, from the correlation amplitudes (\ref{eq6}b), it may be inferred that the exact nature of cross-DFS decoherence free distributions, if any exist, will depend considerably on the particular {\it environment interaction algebra} generated by $H_B$, the couplings $B_k$ and the identity $I_B$ on $\cal B$. Foregoing a comprehensive analysis, we aim to prove that such distributions indeed exist in the familiar setting of a harmonic [bosonic] environment with linear couplings. Then Hamiltonian (\ref{eq2}) becomes, in the usual notation, 

\begin{mathletters}
\label{eq7}

\[
H = H_R + \sum\limits_q {\hbar \omega_q b^{\dagger}_q b_q } + \sum\limits_k {R_k \otimes \sum\limits_q { \left({ \chi_{kq} b^{\dagger}_q +\chi_{kq}^* b_q }\right) } } = 
\]

\begin{equation}
= H_R + \sum\limits_q {\hbar \omega_q b^{\dagger}_q b_q } + \sum\limits_{J,q} { P_J \otimes \left({ \mu_{Jq} b^{\dagger}_q +\mu_{Jq}^* b_q }\right) }\;\;.
\end{equation}

\noindent where in the last line above the coupling constants have been  redefined for convenience as

\begin{equation}
\mu_{Jq} = \sum\limits_k \mu_{k,J} \chi_{kq}\;\;.
\end{equation}

\end{mathletters}

\noindent In this case, each DFS creates a decoupled environment of displaced harmonic modes, driven by Hamiltonians of the form

\begin{mathletters}
\label{eq7a1}

\begin{equation}
H_{B,J} \equiv \sum\limits_q {\hbar \omega_q b^{\dagger}_q b_q } + \sum\limits_q { \left({ \mu_{Jq} b^{\dagger}_q +\mu_{Jq}^* b_q }\right) } = \sum\limits_q {\hbar \omega_q b^{\dagger}_{Jq} b_{Jq} } - \Omega_J^0\;\;,
\end{equation}

\noindent with statically displaced modes described by

\begin{equation}
b_{Jq} = b_q +\frac{\mu_{Jq}}{\hbar\omega_q}\;\;,
\end{equation}

\noindent and energy shifts given by

\begin{equation}
\Omega_J^0 = \sum\limits_q \frac{|\mu_{Jq}|^2}{\hbar\omega_q}\;\;.
\end{equation}

\end{mathletters}

\noindent Accordingly, decomposition (\ref{eq4}) for the total propagator becomes

\begin{equation}
\label{eq7a2}
e^{-\frac{i}{\hbar}Ht} = e^{-\frac{i}{\hbar}H_Rt} \sum\limits_J { e^{\frac{i}{\hbar}\Omega_J^0 t}\exp\left[{-\frac{i}{\hbar}\left(\sum\limits_q \hbar\omega_q b^{\dagger}_{Jq} b_{Jq} \right)t}\right] \otimes P_J  }  \;\;.
\end{equation}

\subsection{Simple examples of multi-DFS decoherence free states: pure coherent-product states}

Given the structure of the propagator (\ref{eq7a2}), it is not difficult to identify {\it pure} states of the joint register-environment system that generate nontrivial, multi-DFS decoherence free solutions. Consider, for instance, a total pure state $|\Phi \rangle$ formed as a superposition of unnormalized [but orthogonal] register states $| \phi^R_J \rangle$ defined on [some] individual DFS, each correlated with the displaced vacuum $|0_J \rangle$ of the corresponding environment Hamiltonian $H_{B,J}$ [Eq.(\ref{eq7a1}a)], i.e.,

\begin{equation}
\label{eq7a3}
|\Phi \rangle = \sum\limits_{J} {| \phi^R_J \rangle \otimes |0_J \rangle} \;\;.
\end{equation}

\noindent Its time evolution follows straightforwardly by applying expression (\ref{eq7a2}) for the total propagator, and reads

\begin{equation}
\label{eq7a4}
|\Phi \rangle = \sum\limits_{J} {e^{-\frac{i}{\hbar}H_Rt}e^{\frac{i}{\hbar}\Omega_J^0 t} | \phi^R_J \rangle \otimes |0_J \rangle} \;\;.
\end{equation}

\noindent Obviously, the associated register reduced state, given by the reduced density matrix

\begin{equation}
\label{eq7a5}
\hat\rho^R(t) = e^{-\frac{i}{\hbar}H_Rt} \cdot \left[{ \sum\limits_{J, J'}{e^{\frac{i}{\hbar}\left( \Omega_J^0 - \Omega_{J'}^0 \right)t} | \phi^R_J \rangle \langle 0_{J'} | 0_J \rangle \langle \phi^R_{J'}|  } }\right] \cdot e^{\frac{i}{\hbar}H_Rt}\;\;,
\end{equation}

\noindent evolves unitarily. Moreover, it evolves in an unperturbed, decoherence free manner, provided the energy shifts for the DFS involved are identical, that is

\begin{mathletters}
\label{eq7a6}

\begin{equation}
\Omega_J^0 - \Omega_{J'}^0 \equiv \sum\limits_q \frac{|\mu_{Jq}|^2}{\hbar\omega_q} - \sum\limits_q \frac{|\mu_{J'q}|^2}{\hbar\omega_q} = 0\;\;.
\end{equation}

\noindent It also involves nonzero correlations between register states belonging to distinct DFS if the displaced environment vacua associated to those DFS have nonvanishing overlap, such that

\begin{equation}
\langle 0_{J'} | 0_J \rangle \neq 0\;\;.
\end{equation}

\end{mathletters}

\noindent One may recall that reciprocally displaced harmonic vacua generate unitarily equivalent Fock spaces iff their overlap amplitude is nonvanishing. Hence condition (\ref{eq7a6}b) requires in effect that the corresponding environment vacua must reside in {\it unitarily equivalent spaces}. That is, e.g., the Fock basis states $ |\{n_{J'} \}\rangle$ generated from $ | 0_{J'} \rangle $ must reside in the Fock space spanned by the states $ |\{ n_J \}\rangle$ generated from $\{ | 0_J \rangle \}$, and can be related to the latter by a unitary transformation that is well-defined on the subtended space, as $|\{n_{J'}\}\rangle = U(b_{Jq}, b_{Jq}^\dagger) |\{ n_J \}\rangle$ [in particular, $|0_{J'}\rangle = U(b_{Jq}, b_{Jq}^\dagger) |0_J \rangle$]. If the vacua are unitarily inequivalent, such that $\langle 0_{J'} | 0_J \rangle= 0$, the reduced state (\ref{eq7a5}) does evolve unitarily, but as a trivial, separable superposition of distributions belonging to orthogonal DFS. 

A more sophisticated example is obtained if the DFS vacua $| 0_J \rangle$ are replaced in the total state (\ref{eq7a3}) by arbitrary coherent states of the corresponding displaced modes, defined by

\[
b_{Jq} |\{ \beta_{J,q} \} \rangle = \beta_{J,q}  |\{ \beta_{J,q} \} \rangle\;\;.
\]

\noindent In addition, let us allow for the possibility that every orthonormal register basis state $|\phi^R_{J\nu} \rangle$ be associated with a different coherent state. The result is a Davydov coherent-product state 

\begin{equation}
\label{eq7a7}
|\Phi \rangle = \sum\limits_{J\nu} {c_{J\nu} | \phi^R_{J\nu} \rangle \otimes |\{\beta_{J\nu, q} \} \rangle}\;\;,
\end{equation}   

\noindent which propagates as

\begin{equation}
\label{eq7a8}
|\Phi \rangle = \sum\limits_{J\nu} {c_{J\nu} e^{-\frac{i}{\hbar}H_Rt}e^{\frac{i}{\hbar}\Omega_J^0 t} | \phi^R_{J\nu} \rangle \otimes |\{\beta_{J\nu, q}(t) \} \rangle} \;\;.
\end{equation}

\noindent The states $|\{\beta_{J\nu, q}(t) \} \rangle$ are coherent states evolved under the DFS-specific environment Hamiltonians $H_{B,J}$, and are determined, as usual, by 

\[
\beta_{J\nu, q}(t) = \beta_{J\nu, q} e^{-i\omega_q t}\;\,.
\]

\noindent The reduced register state reads now

\begin{equation}
\label{eq7a9}
\hat\rho^R(t) = e^{-\frac{i}{\hbar}H_Rt} \cdot \left[{ \sum\limits_{J\nu, J'\nu'}{e^{\frac{i}{\hbar}\left( \Omega_J^0 - \Omega_{J'}^0 \right)t} c_{J\nu} c_{J'\nu'}^* | \phi^R_{J\nu} \rangle \langle \beta_{J'\nu', q}(t) | \beta_{J\nu, q}(t) \rangle \langle \phi^R_{J'\nu'}|  } }\right] \cdot e^{\frac{i}{\hbar}H_Rt}\;\;,
\end{equation}

\noindent and the corresponding conditions for nontrivial decoherence free propagation amount to

\noindent i)  the energy-shift condition (\ref{eq7a6}a); 

\noindent ii) the requirement that the interference amplitudes for coherent states belonging to different DFS be stationary in time,

\[
\frac{d}{dt}\langle \beta_{J'\nu', q}(t) | \beta_{J\nu, q}(t) \rangle =0 \;\; ;
\]

\noindent iii) the requirement that [some] coherent states $|\beta_{J\nu, q}(t) \rangle$ associated to different DFS reside in unitarily equivalent Fock spaces, such that 

\[
\langle \beta_{J'\nu', q}(t) | \beta_{J\nu, q}(t) \rangle \neq 0\;\;.
\]

\noindent The second condition becomes more intuitive if recalled that the coherent states $| \beta_{J\nu, q}(t) \rangle$ and $| \beta_{J'\nu', q}(t) \rangle$ move on "orbits" centered on the reciprocally displaced vacua $|0_J\rangle$ and $|0_{J'}\rangle$. Assuming condition (ii) satisfied, condition (iii) requires then that this movement be synchronized or phased so that the overlap $\langle \beta_{J'\nu', q}(t) | \beta_{J\nu, q}(t) \rangle$ remains constant in time. Note that coherent states associated to the same DFS are always synchronized and their overlap, if nonvanishing, is necessarily time-independent.\\

Although an analysis of conditions (ii) and (iii) above is straightforward, we postpone following this direction. Instead, we proceed to construct a class of finite-temperature generalizations of these Davydov states. Subsequently it is shown that these {\it thermal, statistical states} also include a fairly wide set of decoherence free, multi-DFS states. Under this extended perspective, conditions (ii) and (iii) become particular instances of more general requirements for decoherence free propagation, and their discussion is covered under the general case. \\

\subsection{The finite temperature coherent-product states}

The density matrix generalization of the Davydov coherent-product ansatz \cite{mine} that we purport to employ, can be regarded in the current context as arising from a particular set of initial density matrices $\hat\rho(0)$. Following this point of view, we introduce the coherent-product distributions in a self-contained manner [independent of the approach in ref.\cite{mine}], via a formal protocol for the preparation of the initial state. An alternative possibility specific to the spin-boson model is also suggested in Sec.VI.\\

\subsubsection{Preparation of the initial coherent-product state}

Consider thus the following procedure: 

i) Let the register couplings (\ref{eq3}) be modified by external means such that the original structure of the register DFS is refined, while the strength of the coupling eigenvalues is arbitrarily altered. For notational convenience, we assume that the DFS degeneracy is completely lifted, but more realistic cases can be easily substituted by allowing some of the coupling eigenvalues to coalesce at identical values. In other words, if $\{ |\phi^R_{J\nu} \rangle \}$ is an orthonormal register basis yielding an irrep on $\cal R$ for $t>0$, let all DFS be reduced to the 1-dimensional subspaces of the individual states $|\phi^R_{J\nu}\rangle$, with corresponding projectors $P_{J \nu} = |\phi^R_{J\nu}\rangle \langle \phi^R_{J\nu}|$. If $\overline{\mu}_{k,J\nu}$ denote the new coupling eigenvalues, decomposition (\ref{eq3}) for the register couplings becomes 

\[
R_k = \sum\limits_{J,\nu} {\overline{\mu}_{k,J\nu}P_{J\nu} }
\]

\noindent and the corresponding Hamiltonian amounts to

\[
H' = H_R  + H_B  + \sum\limits_{J\nu} { P_{J\nu} \otimes  \left[ {\sum\limits_{k,q} { \overline{\mu}_{k, J\nu} \left({ \chi_{kq} b^{\dagger}_q +\chi_{kq} b_q }\right) } }\right] }\;. 
\]

ii) At some time $t_0 < 0$ the register, in an arbitrary state $\hat\rho^R$, is brought into contact with the uncorrelated environment in thermal equilibrium at temperature $T$, such that

\begin{equation}
\label{eq8}
\hat\rho (t_0) = \hat \rho^R \otimes \hat\rho^B_T\;\;,
\end{equation}

\noindent with

\[
\hat\rho^B_T = \frac{1}{Z_B} \exp\left[{-\frac{H_B}{k_B T}}\right]\;\;.
\]

iii) The total system is subsequently allowed to evolve under the modified Hamiltonian $H'$ above, until time $t=0$ when the original couplings are instantly restored. Using the appropriate version of Eq.(\ref{eq4}), the state so prepared at $t=0$ is easily seen to be

\begin{equation}
\label{eq9}
\hat\rho(0) =  \sum\limits_{J\nu, J'\nu'} { P_{J\nu}V^B_{J\nu} \hat\rho^R \hat\rho_{B,T} \left(V^B_{J'\nu'} \right)^{\dagger}P_{J'\nu'} }  \;\;,
\end{equation}

\noindent where irrelevant phase factors have been absorbed into the register basis states $|\phi^R_{J\nu} \rangle$, and we introduced the short-hand notation 

\[
V^B_{J\nu} = \exp\left[{ -\frac{i}{\hbar}\left( { H_B + \sum\limits_k { \overline{\mu}_{k, J\nu}B_k} }\right)t_0 }\right] =
\]

\begin{equation}
\label{eq10}
= \exp \left[{ -\frac{i}{\hbar} \left({ \sum\limits_q {\hbar \omega_q b^{\dagger}_q b_q } + \sum\limits_q { \hbar \omega_q \left({ \overline\beta_{J\nu,q} b^{\dagger}_q +\overline\beta_{J\nu,q}^* b_q  } \right)} }\right) t_0 }\right]\;\;.
\end{equation}

\noindent Also,  the adimensional parameters $\overline\beta_{J\nu,q}$ in the last line above are defined by 

\[
\hbar\omega_q \overline\beta_{J\nu,q} = \sum\limits_k {\overline\mu_{k,J\nu} \chi_{kq} }\;\;.
\]

\noindent The environment contributions $\left[{ V^B_{J\nu}\hat\rho_{B,T} \left(V^B_{J'\nu'}\right)^{\dagger} }\right]$ to the density matrix (\ref{eq9}) can be brought to a more convenient form if the factors $V^B_{J\nu}$ are written as

\[
V^B_{J\nu} \cdot \exp\left[\frac{i}{\hbar} \left({ \sum\limits_q \hbar\omega_q b_q^{\dagger} b_q }\right) t_0 \right ]= 
\]

\[
=\exp \left[ i\sum\limits_q \omega_q t_0 |\overline\beta_{J\nu,q}|^2 \right] \cdot \exp\left[ -\sum\limits_q \left( \overline\beta_{J\nu,q} b_q^{\dagger} - \overline\beta_{J\nu,q}^* b_q\right) \right]\cdot 
\]

\[
\cdot \exp\left[-\frac{i}{\hbar} \left({ \sum\limits_q \hbar\omega_q b_q^{\dagger} b_q }\right) t_0 \right ]\cdot \exp\left[ \sum\limits_q \left( \overline\beta_{J\nu,q} b_q^{\dagger} - \overline\beta_{J\nu,q}^* b_q\right) \right] \exp\left[\frac{i}{\hbar} \left({ \sum\limits_q \hbar\omega_q b_q^{\dagger} b_q }\right) t_0 \right ]\;\;.
\]
 
\noindent The last three factors on the right hand side can be easily rearranged as $\exp\left[\sum\limits_q \left( \overline\beta_{J\nu,q}e^{i\omega_q t_0} b_q^{\dagger} - \overline\beta_{J\nu,q}^*e^{-i\omega_q t_0} b_q\right) \right]$, to yield 

\begin{equation}
\label{eq11}
V^B_{J\nu} \cdot \exp\left[\frac{i}{\hbar} \left({ \sum\limits_q \hbar\omega_q b_q^{\dagger} b_q }\right) t_0 \right ]= \exp \left[ i\theta_{J\nu} \right] \cdot \exp\left[ \sum\limits_q \left( \beta_{J\nu,q}(0) b_q^{\dagger} - \beta_{J\nu,q}^*(0) b_q\right) \right] \;\;,
\end{equation} 

\noindent where the displacement parameters $\beta_{J\nu,q}(0) = \overline\beta_{J\nu,q} \left( e^{i\omega_q t_0} -1\right) $ can be retained as final parametrization variables. Use of identity (\ref{eq11}) in expression (\ref{eq9}) now leads to the following defining expression for the sought class of initial density matrices:

\begin{equation}
\label{eq12}
\hat\rho(0) =  \sum\limits_{J\nu, J'\nu'} { P_{J\nu} \exp\left[ \sum\limits_q \left( \beta_{J\nu,q}(0) b_q^{\dagger} - \beta_{J\nu,q}^*(0) b_q\right) \right] \hat\rho^R \hat\rho^B_T \exp\left[ -\sum\limits_q \left( \beta_{J'\nu',q}(0) b_q^{\dagger} - \beta_{J'\nu',q}^*(0) b_q\right) \right] P_{J'\nu'} }
\end{equation}

\noindent where the phase factors $\theta_{J\nu}$ are absorbed again into the corresponding register states. 

The density matrix (\ref{eq12}) describes a register entangled with an environment in a superposition of Gaussian states, as shown by the environment reduced density matrix 

\[
\hat\rho^B(0) = Tr_R[\hat\rho(0)]=\sum\limits_{J\nu} { \frac{Tr_R(P_{J\nu}\hat\rho^R)}{Z_B} \exp\left[{-\frac{1}{k_B T} \sum\limits_q {\hbar\omega_q (b_q^{\dagger}-\beta_{J\nu}^*(0))(b_q -\beta_{J\nu}(0)) } }\right] }\;\;.
\]

\noindent In the particular case when all displacements $\beta_{J\nu}(0)$ vanish, or alternatively, when the preparation procedure is skipped ($t_0=0$), the environment remains in a thermal, uncorrelated state. Another interesting limit arises when the environment is cooled at zero temperature, such that $\hat\rho^B_T=|0_B\rangle \langle 0_B|$, and the register state $\hat\rho^R$ corresponds to a pure state, $\hat\rho^R = |\Phi^R\rangle \langle \Phi^R | $. Then the total state (\ref{eq12}) also describes a pure state, of the form

\[
|\Psi \rangle = \sum\limits_{J\nu} { c_{J\nu} |\phi^R_{J\nu} \rangle \otimes |\{ \beta_{J\nu}(0) \}\rangle   }\;\;,
\]

\noindent where the coefficients are given by $c_{J\nu} =\langle \phi^R_{J\nu} |\Phi^R \rangle$, $\sum\limits{|c_{J\nu}|^2} = 1$, and the environment factors read $ |\{ \beta_{J\nu}(0) \}\rangle = \exp\left[ \sum\limits_q \left( \beta_{J\nu,q}(0) b_q^{\dagger} - \beta_{J\nu,q}^*(0) b_q\right) \right] |0_B\rangle$. One recognizes without difficulty a coherent-product Davydov state. Interestingly enough, the general expression (\ref{eq12}) can be brought to a form reminiscent of the Davydov ansatz as well, by the formal artifice of square-root factorization. Indeed, if we introduce a square-root representation for the density matrices $\hat\rho^R$ and $\hat\rho^B_T$, i.e.,

\[
\hat\rho^R = \hat\gamma^R\cdot (\hat\gamma^R)^{\dagger}\;\;
\] 

\[
\hat\rho^B_T = \hat\gamma^B_T \cdot (\hat\gamma^B_T)^{\dagger}\;\;,
\]

\noindent the entire density (\ref{eq12}) is factorized as

\[
\hat\rho(0) = \hat\gamma(0) \cdot \hat\gamma^{\dagger}(0)\;\;,
\]  

\noindent with 

\[
\hat\gamma(0) = \sum\limits_{J\nu} { \hat\gamma^R_{J\nu}(0) \otimes \hat\gamma^B_{J\nu}(0) }\;\;.
\]

\noindent The register and environment factors in the decomposition above now read respectively

\[
\hat\gamma^R_{J\nu}(0) = P_{J\nu} \hat\gamma^R\;\;,
\]

\noindent and

\[
\hat\gamma^B_{J\nu} (0) = \exp\left[ \sum\limits_q \left( \beta_{J\nu,q}(0) b_q^{\dagger} - \beta_{J\nu,q}^*(0) b_q\right) \right] \cdot \hat\gamma^B_T\;\;.
\]

\noindent The total density matrix $\hat\rho(0)$ acquires in this way a structure similar to the density matrix for a Davydov pure state, where the various pure state factors are replaced by [nonhermitian] operator factors. Moreover, the register factors are orthonormal in the sense of the usual trace inner product on the space of linear operators on $\cal R$, i.e., $(\hat\gamma^R_{J\nu}(0)|\hat\gamma^R_{J'\nu'}(0)) \equiv Tr_R[\hat\gamma^R_{J\nu}(0)\cdot (\hat\gamma^R_{J'\nu'}(0))^{\dagger} ] = \delta_{J\nu,J'\nu'}$, and the environment factors $\hat\gamma^B_{J\nu} (0)$ are easily verified to be square roots of Gaussian distributions of the form $\exp\left[{-\frac{1}{k_B T} \sum\limits_q {\hbar\omega_q (b_q^{\dagger}-\beta_{J\nu}^*(0))(b_q -\beta_{J\nu}(0)) } }\right]$. This formal analogy prompts the designation of states described by Eq.(\ref{eq12}) as density matrix coherent-product states. It was found in ref.\cite{mine}, and it is shown shortly for the case at hand, that such states preserve their form throughout an evolution driven by a Hamiltonian of type (\ref{eq7}), provided the register states are restricted to a direct sum of DFS. 

Before proceeding in this direction, let us stress that the register density matrix $\hat\rho^R$ in expression (\ref{eq12}) above {\it is not identical} to the initial reduced density matrix $\hat \rho^R(0)$ unless all displacements $\beta_{J\nu,q}(0)$ vanish, and the total density matrix reduces to the thermal, uncorrelated form (\ref{eq8}). In all other cases, the correct reduced density matrix at $t=0$ is entangled with the environment, and reads 

\[
\hat\rho^R(0) =  \sum\limits_{J\nu, J'\nu'} {\eta_{J\nu,J'\nu'}(0) P_{J\nu}\hat\rho^R P_{J'\nu'} }\;\;,
\]

\noindent where

\[
\eta_{J\nu,J'\nu'}(0) = Tr_B\left[ {\exp\left[ \sum\limits_q \left( \beta_{J\nu,q}(0) b_q^{\dagger} - \beta_{J\nu,q}^*(0) b_q\right) \right]  \hat\rho_{B,T}\exp\left[ -\sum\limits_q \left( \beta_{J'\nu',q}(0) b_q^{\dagger} - \beta_{J'\nu',q}^*(0) b_q\right) \right]  }\right]\;\;.
\]

\noindent Thus the matrix $\hat\rho^R$ represents only a register parametrization variable for $\hat\rho(0)$, just as the displacements $\beta_{J\nu,q}(0) $ parametrize the environment contribution. More precisely, the correlation amplitudes between the register states $|\phi^R_{J\nu}\rangle$ projected by $P_{J\nu}$ are given as

\[
\langle \phi^R_{J\nu} | \hat\rho^R(0) | \phi^R_{J\nu}\rangle = \eta_{J\nu,J'\nu'}(0) \langle \phi^R_{J\nu} | \hat\rho^R| \phi^R_{J\nu}\rangle
\]  

\noindent and include both an environment-mediated component contributed by $\eta_{J\nu,J'\nu'}(0)$ and a parametric register component represented by $\langle \phi^R_{J\nu} | \hat\rho^R| \phi^R_{J\nu}\rangle$.\\

\subsubsection {The time-dependent coherent-product solution}

The density matrix evolved under Hamiltonian (\ref{eq7}) from initial density matrix (\ref{eq12}) can be calculated straightforwardly with expression (\ref{eq4}) for the evolution operator. After a rearrangement of the type leading to identity (\ref{eq11}), it is found to read

\begin{equation}
\label{eq13}
\hat\rho(t) =  \exp\left[{ -\frac{i}{\hbar}H_Rt }\right] \left[{ \sum\limits_{J\nu, J'\nu'} { e^{i \Theta_{J\nu}(t)} P_{J\nu}  U^B_{J\nu}(t)  \hat\rho^R \hat\rho^B_T \left( U^B_{J'\nu'}(t)\right)^{\dagger}} P_{J'\nu'} e^{ -i \Theta_{J'\nu'}(t)} }\right]  \exp\left[{ \frac{i}{\hbar}H_Rt }\right]
\end{equation}

\noindent where the phase factors $\Theta_{J\nu}(t)$ can be shown to amount to

\begin{mathletters}
\label{eq14}

\begin{equation}
\Theta_{J\nu}(t) = -\frac{1}{2\hbar} \int\limits_0^t d\tau \sum\limits_q { \left({ \mu_{Jq}^*\beta_{J\nu,q}(\tau) + \mu_{Jq}\beta_{J\nu,q}^*(\tau) }\right) }\;\;,
\end{equation}

\noindent and $U^B_{J\nu}(t)$ are environment displacement operators of the form

\begin{equation}
U^B_{J\nu}(t) = \exp\left[ \sum\limits_q \left( \beta_{J\nu,q}(t) b_q^{\dagger} - \beta_{J\nu,q}^*(t) b_q\right) \right] \;\;.
\end{equation}

\end{mathletters}

\noindent In Eqs.(\ref{eq14}) above, the environment displacements $\beta_{J\nu,q}(t)$ evolve as

\begin{equation}
\label{eq15}
\beta_{J\nu,q}(t) = \left( { \beta_{J\nu,q}(0) + \frac{\mu_{Jq}}{\hbar\omega_q} }\right) e^{-i\omega_q t}- \frac{\mu_{Jq}}{\hbar\omega_q} \;\;.
\end{equation}

\noindent As anticipated, solution (\ref{eq13}) indeed preserves the coherent-product character of the initial condition (\ref{eq12}). In a square-root decomposition $\hat\rho(t) =\hat\gamma(t) \cdot \hat\gamma^{\dagger}(t)$, a representative [nonhermitian] square-root $\hat\gamma(t)$ takes the form 

\[
\hat\gamma(t) = \sum\limits_{J\nu} { \hat\gamma^R_{J\nu}(t) \otimes \hat\gamma^B_{J\nu}(t) }\;\;,
\]

\noindent with factors given by

\[
\hat\gamma^R_{J\nu}(t) = e^{i\Theta_{J\nu}(t)} \exp\left[{ -\frac{i}{\hbar}H_Rt }\right] \hat\gamma^R_{J\nu}(0)\;\;,
\]

\[
(\hat\gamma^R_{J\nu}(t)|\hat\gamma^R_{J'\nu'}(t)) \equiv Tr_R[\hat\gamma^R_{J\nu}(t)\cdot (\hat\gamma^R_{J'\nu'}(t))^{\dagger} ] = \delta_{J\nu,J'\nu'}\;\;, 
\]

\noindent and

\[
\hat\gamma^B_{J\nu} (t) = \exp\left[ \sum\limits_q \left( \beta_{J\nu,q}(t) b_q^{\dagger} - \beta_{J\nu,q}^*(t) b_q\right) \right] \cdot \hat\gamma^B_T\;\;.
\]

\noindent  Obviously, the environment remains in a nonstationary statistical superposition of Gaussian states, with a reduced state

\[
\hat\rho^B(t) = Tr_R[\hat\rho(t)]=\sum\limits_{J\nu} { \frac{Tr_R(P_{J\nu}\hat\rho^R)}{Z_B} \exp\left[{-\frac{1}{k_B T} \sum\limits_q {\hbar\omega_q (b_q^{\dagger}-\beta_{J\nu}^*(t))(b_q -\beta_{J\nu}(t)) } }\right] }\;\;.
\]

\noindent It is worth noting in particular that Eq.(\ref{eq13}) offers a closed density matrix form for the well-known exact solution to the decoherence problem with an uncorrelated, thermal initial state of type (\ref{eq8}) \cite{Cal-L}. From the above observation on the state of the environment, we obtain as a direct corollary that {\it the evolution of an initial state (\ref{eq8}) under Hamiltonian (\ref{eq7}) drives the environment into a nonequilibrium state of type (\ref{eq13}), strongly entangled with the register, unless the initial register state is restricted to a single DFS}. Conversely, and with reference also to the preparation procedure described earlier, the coherent-product solutions can be regarded as straightforward generalizations of the standard, thermal solution. The formal essence of this generalization is seen in expression (\ref{eq15}) for the time-dependence of the environment displacements $\beta_{J\nu,q}(t)$. That is, the coherent product form (\ref{eq13}) for a generalized initial state (\ref{eq12}) differs from the coherent product form for the thermal solution solely by the presence of nonzero initial displacements $\beta_{J\nu,q}(0)$. It can be said therefore that {\it the standard solution is generalized here by allowing for nonzero initial environment displacements in its coherent-product expression}.

Let us point out also that in the zero-temperature, pure state limit, corresponding as before to $\hat \rho^R = |\Phi^R\rangle \langle \Phi^R|$ and $\hat\rho^B_{T=0} = |0_B\rangle \langle 0_B|$, the coherent-product density matrix (\ref{eq13}) describes a Davydov-type pure state $|\Phi(t) \rangle$ evolving as

\[
|\Phi(t)\rangle = \exp\left[{ -\frac{i}{\hbar}H_Rt }\right] \sum\limits_{J,\nu} {c_{J\nu}e^{i\Theta_{J\nu}(t)}|\phi^R_{J\nu}\rangle \otimes |\beta_{J\nu}(t)\rangle  }\;\;,
\]

\noindent with $c_{J\nu} =\langle \phi^R_{J\nu} |\Phi^R \rangle$, $\sum\limits{|c_{J\nu}|^2} = 1$, and the environment coherent states $|\beta_{J\nu}(t)\rangle$ defined by $ |\beta_{J\nu}(t)\rangle = \exp\left[ \sum\limits_q \left( \beta_{J\nu,q}(t) b_q^{\dagger} - \beta_{J\nu,q}^*(t) b_q\right) \right] |0_B\rangle$. As must be expected, we recover the Davydov state example (\ref{eq7a8}) of Sec.IIIB. The apparently different phase factors are due to the representation of the environment coherent states in terms of unperturbed modes $b_q$, rather than displaced, DFS-specific modes $b_{Jq}$.\\ \\

In the next Section we show that the class of thermal coherent-product states can generate multi-DFS decoherence free reduced register distributions, under reasonably relaxed constraints on the coupling constants $\mu_{Jq}$ and on the symmetry of the environment. The general form of the reduced register state reads now [compare to Eq.(\ref{eq6})]

\begin{equation}
\label{eq16}
\hat\rho^R(t) =  Tr_B[\hat\rho(t)] = \exp\left[{ -\frac{i}{\hbar}H_Rt }\right] \cdot \left[{ \sum\limits_{J\nu, J'\nu'} {\eta_{J\nu,J'\nu'}(t) P_{J\nu}\hat\rho^R P_{J'\nu'} } }\right] \cdot \exp\left[{ \frac{i}{\hbar}H_Rt }\right]\;\;,
\end{equation} 

\noindent and shows that the effect of the environment is conveniently concentrated in the bath-mediated correlation factors 

\begin{equation}
\label{eq17}
\eta _{J\nu, J'\nu' } \left( t \right) = e^{i [\Theta_{J\nu}(t)-\Theta_{J'\nu'}(t)] } Tr_B \left[ {U^B_{J\nu}(t)\hat\rho_{B,T} \left( U^B_{J'\nu'}(t)\right)^{\dagger}} \right] \;\;,
\end{equation}

\noindent where the phase factors $\Theta_{J\nu}$ and the unitary displacement transformations $U^B_{J\nu}$ are defined by Eqs.(\ref{eq14}). The trace in Eq. (\ref{eq17}) can be readily calculated, either in the usual manner, via the symmetric order generating functional for the harmonic oscillator \cite{decoh}, or by the techniques of thermofield dynamics \cite{mine}. It is also helpful to use the prior observation that expression (\ref{eq17}) differs from the corresponding result for uncorrelated, thermal initial conditions only by the specific time-dependence of the displacements $\beta_{J\nu,q}$. The ensuing expression for the bath correlation $\eta _{J\nu, J'\nu' }$ is of the form 

\begin{equation}
\label{eq18a}
\eta _{J\nu, J'\nu' } \left( t \right) = e^{ i[\Theta_{J\nu}(t)-\Theta_{J'\nu'}(t)-\Phi _{J\nu, J'\nu' }(t)] } e^{ - \Gamma _{J\nu, J'\nu' }(t)}\;\;,
\end{equation}

\noindent where the phase factor $\Phi _{J\nu, J'\nu' }$ amounts to

\begin{mathletters}
\label{eq19a}

\begin{equation}
\Phi _{J\nu, J'\nu' } \left( t \right) = \frac{i}{2}\sum\limits_q {\left[ {\beta _{J\nu,q} \left( t \right)\beta _{J'\nu',q }^* \left( t \right) - \beta _{J\nu,q}^* \left( t \right)\beta _{J'\nu',q } \left( t \right)} \right]} \;,
\end{equation}

\noindent while the dissipative factor $\Gamma _{J\nu, J'\nu' }$ is given by

\begin{equation}
\Gamma _{J\nu, J'\nu' } \left( t \right) = \frac{1}{2}\sum\limits_q {\left| {\beta _{J\nu,q
} \left( t \right) - \beta _{J'\nu',q } \left( t \right)} \right|} ^2 \coth \left( {\frac{{\hbar \omega _q }}{{2 k_B T}}} \right) \;\;.
\end{equation}

\end{mathletters}

\noindent For the spin-boson model (\ref{eq1}) with an uncorrelated, thermal initial condition, it can be verified starightforwardly that these general forms reduce to the expressions recently derived in ref.\cite{decoh1}. 

The physical meaning of the bath-correlation $\eta_{J\nu, J'\nu'}$ becomes apparent when the trace in Eq.(\ref{eq17}) is expressed as a thermal average, to the result that

\begin{equation}
\label{eq20a}
\eta _{J\nu, J'\nu' } \left( t \right) = e^{i [\Theta_{J\nu}(t)-\Theta_{J'\nu'}(t)] } \sum \limits_{ \{n_q\} } { \frac{\exp \left({ -E_{ \{n_q\} }/k_B T }\right) }{Z} \langle \{ n_{q, J\nu}(t) \} | \{ n_{q, J'\nu'}(t) \} \rangle } \;,
\end{equation}

\noindent where 

\[
| \{n_{q, J\nu}(t) \} \rangle = \exp \left[ { \sum\limits_{q} { \left( { \beta_{J\nu,q}(t) b_q^{\dagger} - \beta_{J\nu,q}^*(t) b_q } \right) } }\right] | \{n_q \} \rangle \;\; ,
\]

\noindent with $| \{n_q \} \rangle $ the excited bath state with $n_q$ quanta in mode $q$, and $E_{\{n_q\} }=\sum\limits_q{\hbar\omega_q n_q}$ the excitation energy of the state $| \{n_q \} \rangle $. Recall that the states $| \{n_{q,J\nu}(t) \} \rangle$ can be understood in terms of the displaced modes 

\[
b_{q,J\nu}^{\dagger}(t) \equiv \exp \left[ { \sum\limits_{q} { \left( { \beta_{J\nu,q}(t) b_q^{\dagger} - \beta_{J\nu,q}^*(t) b_q } \right) } }\right] \cdot b_q^{\dagger} \cdot \exp \left[ - { \sum\limits_{q} { \left( { \beta_{J\nu,q}(t) b_q^{\dagger} - \beta_{J\nu,q}^*(t) b_q } \right) } }\right] = b_q^{\dagger}-\beta_{J\nu,q}^*(t)
\]

\noindent as [time-dependent] environment states with $n_q$ displaced quanta of mode $q$ excited over a displaced vacuum

\[
| \{\beta_{q,J\nu}(t)\} \rangle \equiv \exp \left[ { \sum\limits_{q} { \left( { \beta_{J\nu,q}(t) b_q^{\dagger} - \beta_{J\nu,q}^*(t) b_q } \right) } }\right] | 0 \rangle  \;\;.
\]

\noindent That is, 

\[
|\{n_{q,J\nu}(t) \} \rangle= \prod\limits_q {\frac{1}{\sqrt{n_q!} } \left({b_{q,J\nu}^{\dagger}(t)}\right)^{n_q} | \{\beta_{q,J\nu}(t)\} \rangle }\;\;.
\]

\noindent Hence the bath factors $\eta _{J\nu, J'\nu' } \left( t \right)$ represent, up to a phase factor, the thermally weighted sum of interference amplitudes between the time-dependent coherent environment states entangled with the register states $|\phi^R_{J\nu} \rangle$ and $|\phi^R_{J'\nu'} \rangle$ [i.e., with the register factor $P_{J\nu} \hat\rho_R \cdot P_{J'\nu'} $] in the overall density matrix. \\ \\

\section{Decoherence free states entangled with a nonequilibrium environment : general conditions in the coherent-product ansatz}

A first hint that nontrivial, multi-DFS decoherence free states are possible in the thermal coherent-product ansatz, comes from the straightforward observation that the harmonic time-dependence law for the environment displacements given by Eq.(\ref{eq15}) admits the stationary points 

\[
\overline\beta_{J\nu,q} = -\frac{\mu_{Jq}}{\hbar\omega_q} \;\;.
\]

\noindent If the initial state is such that the environment modes are displaced exactly over these positions, the reduced environment state remains stationary and the bath-correlation factors in expression (\ref{eq16}) for the register reduced state varies only through the phase factors $\Theta_{J\nu}$. The origin of this effect can be seen by noting that in this case the action of the displacement operators (\ref{eq14}b) on the excited states $|\{ n_q \}\rangle$ of the unperturbed environment generates, up to a phase factor, the similarly excited states of the corresponding DFS-specific Hamiltonian, i.e., 

\[
U^B_{J\nu}|\{ n_q \}\rangle \to \exp\left[ -\sum\limits_q{ \left( \frac{\mu_{Jq}}{\hbar\omega_q}b^{\dagger}_q - \frac{\mu_{Jq}^*}{\hbar\omega_q}b_q \right) } \right] |\{ n_q \}\rangle = |\{ n_{J,q} \}\rangle
\]

\noindent where [see Eqs.(\ref{eq7a1})]

\[
H_{B,J} |\{ n_{J,q} \}\rangle = \left( \sum\limits_q { \hbar\omega_q n_q } - \Omega_J^0 \right)|\{ n_{J,q} \}\rangle
\]

\noindent Taking this into the environment factors $\left[ {U^B_{J\nu}(t)\hat\rho_{B,T} \left( U^B_{J'\nu'}(t)\right)^{\dagger}} \right]$ of the coherent-product density matrix (\ref{eq13}), shows that the register states are actually correlated with eigenstates of the environment Hamiltonians $H_{B,J}$, which vary only through phase factors under the action of the total propagator (\ref{eq7a2}). 

Returning now to the bath-correlation factors in the register reduced state (\ref{eq16}), observe that the remaining phase factors $\Theta_{J\nu}$ become proportional to the energy-shifts (\ref{eq7a1}c) and linear in time, as 

\[
\Theta_{J\nu}(t) = \frac{t}{\hbar} \sum\limits_q { \frac{ \left|{ \mu_{Jq}}\right|^2}{\hbar\omega_q} } \equiv \frac{1}{\hbar}\Omega_J^0 t \;\;.
\]

\noindent Consequently, the evolution of the register state (\ref{eq16}) acquires the unitary, quasi-unperturbed form

\[
\hat\rho^R(t) =   \exp\left[{ -\frac{i}{\hbar}H_R\;t }\right] \cdot \left[{ \sum\limits_{J\nu, J'\nu'} { e^{\frac{i}{\hbar}(\Omega_J^0-\Omega_{J'}^0) t}\; \overline \eta_{J\nu,J'\nu'} P_{J\nu}\hat\rho^R P_{J'\nu'} } }\right] \cdot \exp\left[{ \frac{i}{\hbar} H_R t }\right] = 
\]

\[
= \exp\left[{ -\frac{i}{\hbar}(H_R-\sum\limits_J \Omega_J^0 P_J)\;t }\right] \cdot \left[{ \sum\limits_{J\nu, J'\nu'} {  \overline \eta_{J\nu,J'\nu'} P_{J\nu}\hat\rho^R P_{J'\nu'} } }\right] \cdot \exp\left[{ \frac{i}{\hbar}(H_R-\sum\limits_J \Omega_J^0 P_J)\;t }\right]\;\;.
\]

\noindent As in the first pure state example of Sec.IIIB, an exact and nontrivial unperturbed propagation requires that the reference energies of the contributing DFSs be identical, such that

\[
\Omega_J^0-\Omega_{J'}^0 \equiv \sum\limits_q { \frac{ \left|{ \mu_{Jq}}\right|^2}{\hbar\omega_q} } - \sum\limits_q { \frac{ \left|{ \mu_{J'q}}\right|^2}{\hbar\omega_q} } = 0\;\;,
\]

\noindent and that some correlation factors between distinct DFS [$J\neq J'$] be nonzero. Otherwise, as will be shown to happen for model (\ref{eq1}) in an environment with a one-dimensional spectral density, the correlations reduce to $\overline\eta_{J\nu,J'\nu'} = \delta_{J,J'}$ and the presence of the bath is effectively erased in  $\rho _R \left( t \right)$. The latter becomes a trivial block diagonal distribution on disjoint DFS, and individual decoherence-free states become eventually pointer states \cite{einselection}.\\     

Let us now seek the general conditions for the {\it unitary and unperturbed} propagation of the reduced register state in the coherent-product ansatz, starting from the requirement of time-independent bath factors, i.e.,

\begin{equation}
\label{eq18}
\frac{d}{{dt}}\ln \eta _{J\nu,  J'\nu' }  \equiv   i\left({ \frac{d\Theta _{J\nu} }{dt}-\frac{d\Theta _{J'\nu' } }{dt}-\frac{d\Phi _{J\nu, J'\nu' } }{dt} }\right) - \frac{{d\Gamma _{J\nu, J'\nu' } }}{{dt}} = 0 \;.
\end{equation}

\noindent Substitution of the corresponding expressions, Eqs.(\ref{eq14}a) and (\ref{eq19a}), including the explicit time-dependence (\ref{eq15}) for the bath displacements, and a little algebra yields a Fourier-like sum of the form

\begin{equation}
\label{eq19}
i\hbar \frac{d}{{dt}}\ln \eta _{J\nu, J'\nu' }  = F_0  + \frac{1}{2}\sum\limits_q {\left( {F_{q, + }^* e^{i\omega _q t}  + F_{q, - } e^{ - i\omega _q t} } \right)} \;,
\end{equation}
 
\noindent with coefficients

\begin{mathletters}
\label{eq20}

\begin{equation}
F_0  = \sum\limits_q {\left( {\frac{{\left| {\mu_{Jq} } \right|^2 }}{{\hbar \omega _q }} - \frac{{\left| {\mu_{J'q} } \right|^2 }}{{\hbar \omega _q }}} \right)} \;,
\end{equation}

\begin{equation}
F_{q, + }  = \left( {\mu_{Jq}^*  \beta_{J\nu,q}^0   - \mu_{J'q}^*  \beta_{J'\nu',q}^0}  \right) + \left( {\mu_{Jq}^* \beta_{J'\nu',q}^0  - \mu_{J'q}^* \beta_{J\nu,q}^0 } \right) - \coth \left({\frac{{\hbar \omega _q }}{{2k_B T}}}\right)\left( {\beta _{J\nu,q}^0 -  \beta_{J'\nu',q}^0} \right) \left( {\mu_{Jq}^*  - \mu_{J'q}^* } \right)\;,
\end{equation}

\noindent and, respectively,

\begin{equation}
F_{q, - }  = \left( {\mu_{Jq}^* \beta_{J\nu,q}^0  - \mu_{J'q}^* \beta_{J'\nu',q}^0 } \right) + \left( {\mu_{Jq}^* \beta _{J'\nu',q }^0 - \mu_{J'q}^* \beta_{J\nu,q}^0 } \right) + \coth \left( {\frac{{\hbar \omega _q }}{{2k_B T}}} \right)\left( {\beta_{J\nu,q}^0  - \beta_{J'\nu',q}^0 } \right) \left( {\mu_{Jq}^*  - \mu_{J'q}^* } \right)\;.
\end{equation}

\end{mathletters}

\noindent Here the labels $J'\nu' $ and $J\nu$ have been dropped for notational economy, and we have denoted

\[
\beta_{J\nu,q}^0  = \beta_{J\nu,q} \left( 0 \right) + \frac{{\mu_{Jq} }}{{\hbar \omega _q }}.
\]

\noindent The Fourier-like nature of the right hand side in Eq.(\ref{eq19}) can be easily exploited under the customary assumption of an d-dimensional environment with a continuum spectrum, $\omega_q \to \omega(q)$. We also assume, without loss of generality, a spherical dispersion relation of the form $\omega(q) = \omega(|q|)$, and an environment volume set to unity. In this case, the sum in Eq.(\ref{eq19}) can be rearranged into a Fourier integral by subsuming contributions from degenerate modes on spheres $S_{\omega_q = \omega}$, to the result that condition (\ref{eq18}) is brought to the form 

\begin{equation}
\label{eq21}
F_0  + \frac{1}{2}\int\limits_0^\infty  {d\omega \frac{d\left| q \right|}{d\omega } \left[ {\left( {\oint\limits_{S_\omega  } {dS_q  F_{ q, + }^*}} \right)e^{i\omega t}  + \left( {\oint\limits_{S_\omega  } {dS_q F_{q, - } }} \right)e^{ - i\omega t} } \right] } = 0 \;.
\end{equation}

\noindent Since the cancellation of a constant term $F_0  \ne 0$ demands

\[
\left( {\oint\limits_{S_\omega  } {dS_q F_{ q, \pm } } } \right) \sim \delta (\omega )
\]

\noindent and, in all likelihood, divergent displacements for the static ground mode $\omega  = 0$, it is seen that an unperturbed evolution of the register requires $F_0  = 0$, i.e.,

\begin{mathletters}
\label{eq23}

\begin{equation}
\int {dq\frac{{\left| {\mu_{Jq} } \right|^2 }}{{\hbar \omega _q }}}  = \int {dq\frac{{\left| {\mu_{J'q} } \right|^2 }}{{\hbar \omega _q }}} \;,
\end{equation}

\noindent  as well as 

\begin{equation}
\oint\limits_{S_\omega  } {dS_q F_{ q, \pm } }  = 0
\end{equation}

\end{mathletters}

\noindent for all $J\nu $ and $J'\nu'$. 

Condition (\ref{eq23}a) is easily recognized as the energy-shift condition (i) [Eq.(\ref{eq7a6})a] for the pure coherent-product states of Sec.IIIB, and also retrieved for the particular example in the beginning of this Section. Conditions (\ref{eq23}b) reduce eventually to the simpler linear system

\begin{equation}
\label{eq24}
\oint\limits_{S_\omega  } {dS_q \beta_{J\nu, q}^0 \left( {\mu_{Jq}^*  - \mu_{J'q}^* } \right) }  = 0\;, \;\forall J,J',\nu\;\;,
\end{equation}

\noindent when noted that according to the explicit expressions (\ref{eq20}) the terms in $\coth \left( {\hbar \omega _q /2k_B T} \right)$ contributed by the dissipative exponent $\Gamma _{J\nu, J'\nu'} $ cancel separately. Since this system is temperature independent, it necessarily applies also to the pure state case of Sec.IIIB. In this limit it is seen to ensure that the evolution of the environment coherent states preserves their overlap. By analogy, in the finite temperature situation it secures the synchronization of the Gaussian environment distributions, so that their thermally averaged interference yields a time-independent bath-correlation factor $\eta_{J\nu, J'\nu'}$. 

Thus the unperturbed evolution of a register distribution in the generalized coherent-product ansatz requires simultaneously: \\

i) {\it identical energy reference points (energy-shifts) for the decoupled environments generated by contributing register DFS [Eqs.(\ref{eq23}a) or (\ref{eq7a6})].}

ii) {\it a specific synchronization (phasing) of the entangled environmental modes, through proper displacements $\beta_{J\nu,q}$ [Eqs.(\ref{eq24})];} \\

\noindent For any register of type (\ref{eq1}) states compatible with conditions (i)-(ii) always exist, because for any DFS characterized by $\mu_{Jq}$ there exists another DFS characterized by $(-\mu_{Jq})$, and related to the former by a reversal of all qubits along direction $z$ [$|\uparrow \rangle_n \to |\downarrow \rangle_n$, $| \downarrow \rangle_n \to |\uparrow \rangle_n $], in every member state. This ensures that condition (i) can be satisfied. Condition (ii) is satisfied at least by the trivial solution to Eqs.(\ref{eq24}), $\beta _{J\nu,q}^0  = \beta _{J'\nu',q}^0  = 0$ for all $q$, which corresponds to the case of stationary displacements discussed in the beginning of this section. Nontrivial solutions are essentially conditioned by the degeneracy of the environmental modes. It is worth noting that the number of distinct DFS contributing to a decoherence-free distribution of the type discussed here is theoretically arbitrary [if finite], unless the environment is 1-dimensional and displays only two-fold degenerate modes [$\omega_{-q}=\omega_q$]. For the latter case, the number of distinct DFS involved cannot exceed 2, since the number of unknown $\beta_{J\nu,q}$-s in system (\ref{eq24}) must exceed the number of constraints. 

However, in addition to conditions (i)-(ii) above, the existence of nontrivial environment-entangled state is essentially limited by the requirement that {\it [some of] the bath correlations $\eta_{J\nu, J'\nu'}$ must be nonvanishing}, which means that the corresponding dissipative factors $\Gamma _{J\nu, J'\nu'}$ must be finite, $\Gamma _{J\nu, J'\nu'}<\infty$. Assuming a nontrivial solution to system (\ref{eq24}) does exist, substitution of the resulting bath displacements in expression (\ref{eq19a}b) yields straightforwardly [the time-dependent terms vanish by conditions (\ref{eq23})]

\begin{equation}
\label{eq25}
\Gamma_{J\nu, J'\nu'}  \propto \frac{1}{2} \int{d\omega \frac{d|q|}{d\omega}\coth \left( {\frac{\hbar \omega }{2k_B T}} \right) \oint \limits_{S_\omega} {dS_q \left[ {\left| {\beta_{J\nu,q}^0  - \beta_{J'\nu',q}^0 } \right|^2  + \left| {\frac{\mu_{Jq}  - \mu_{J'q} }{\hbar \omega }} \right|^2 } \right] } }\;\;. 
\end{equation}  
 
\noindent It follows that $\Gamma_{J\nu, J'\nu'}$ cannot be finite unless its value $\Gamma_{J, J'}^0$ for stationary initial displacements [$\beta_{J\nu,q}^0  = \beta_{J'\nu',q}^0  = 0$] is also finite,

\begin{equation}
\label{eq26}
\Gamma_{J, J'}^0  \propto  \frac{1}{2} \int{d\omega \frac{d|q|}{d\omega} (\hbar \omega)^{-2} \coth \left( {\frac{\hbar \omega }{2k_B T}} \right) \oint \limits_{S_\omega} {dS_q  \left| {\mu_{Jq}  - \mu_{J'q} } \right|^2  } } < \infty\;\;.
\end{equation}  

\noindent  This intrinsic stationary dissipative factor $\Gamma_{J, J'}^0$ is seen to be {\it characteristic of the two contributing DFS}, and not of individual contributing states. It sets an upper limit on the magnitude of the corresponding bath correlation $\eta_{J\nu, J'\nu'}$, and therefore on the amplitude of correlation between the associated register states [see Eq.(\ref{eq16}) for the reduced register state $\hat\rho^R(t)$]. It also increases with the temperature [$\Gamma_{J, J'}^0 \to \infty$ as $T \to \infty$] regardless of the exact density of states or of the form of the coupling constants $\mu_{Jq}$, and gradually shrinks the set of register distributions compatible with unperturbed propagation toward trivial states, block diagonal on DFS [$\eta_{J\nu, J'\nu'}\to 0$ as $T\to\infty$ for $J\neq J'$]. Moreover, the  presence of properly phased coherent oscillations of the bath modes [$ \left| {\beta_{J'\nu',q}^0 - \beta_{J\nu,q}^0 } \right|  \geq 0$] results invariably in decreased bath correlations $\eta_{J\nu, J'\nu'}$ and decreased cross-DFS matrix elements in $\hat\rho^R(t)$. Hence the preservation of an unperturbed register evolution seems to involve a trade-off between the stabilizing action of a nonequilibrium environment and the magnitude of the conserved register correlations. 

On the other hand, the finite or infinite character of $\Gamma_{J, J'}^0$ at finite temperatures does depend on both the density of degenerate bath modes, and the coupling constants for the input DFS. To find the source of this effect, let us examine the thermal average representation (\ref{eq20a}) of the bath correlation $\eta_{J\nu, J'\nu'}$ in the limit situation when Eqs.(\ref{eq24}) are trivialy satisfied for $\beta_{J'\nu',q}^0  = \beta_{J\nu,q}^0  = 0$ [$ \beta_{J\nu,q}  = -(\mu_{Jq}/\hbar\omega_q)$, $\beta_{J'\nu',q}=-(\mu_{J'q}/\hbar\omega_q)$], $\Gamma_{J\nu, J'\nu'} \to \Gamma_{J, J'}=\Gamma_{J, J'}^0$ and $\eta_{J\nu, J'\nu'} \to \eta_{J, J'} \sim \exp[-\Gamma_{J,J'}^0]$. To this end, substitute $ \beta_{J\nu,q} \to \beta_{J,q} = -(\mu_{Jq}/\hbar\omega_q)$ and  $\beta_{J'\nu',q}\to \beta_{J',q} =-(\mu_{J'q}/\hbar\omega_q)$ and write the resulting overlaps $\langle \{ n_{q, J} \} | \{ n_{q, J'} \} \rangle $ in the form

\[
\langle \{ n_{q, J} \} | \{ n_{q, J'} \} \rangle = \langle \{n_q\} | \exp \left[ { \sum\limits_{q} { \left({ \frac{ \mu_{Jq}}{\hbar\omega_q} b_q^{\dagger} - \frac{ \mu_{Jq}^*}{\hbar\omega_q} b_q } \right) } }\right]\exp \left[ -{ \sum\limits_{q} { \left({ \frac{ \mu_{J'q}}{\hbar\omega_q} b_q^{\dagger} - \frac{ \mu_{J'q}^*}{\hbar\omega_q} b_q } \right) } }\right] | \{n_q\} \rangle 
\]

\begin{equation}
\label{eq27}
\propto \alpha_{J,J'} \prod\limits_q {\left \langle{ 0_q\left| { b_q^n \exp \left[{- \frac{\mu_{Jq}^*-\mu_{J'q}^*}{\hbar\omega_q} b_q }\right]  \exp \left[{ \frac{\mu_{Jq}-\mu_{J'q}}{\hbar\omega_q} b_q^{\dagger} }\right] \left( b_q^{\dagger}\right )^n }\right| 0_q }\right  \rangle }\;\;,
\end{equation}

\noindent where $\alpha_{J,J'} = \exp\left[{ (1/2)\sum\limits_q {\left ( {\mu_{Jq}^*\mu_{J'q}-\mu_{Jq}\mu_{J'q}^*}\right)/(\hbar\omega_q)^2} }\right] \exp\left[{ (1/2) \sum\limits_q { |\mu_{Jq}-\mu_{J'q}|^2/(\hbar\omega_q)^2} }\right]$. The vacuum averages in the latter expression can be calculated as  

\[
(-1)^n \frac{\partial^n}{\partial \lambda^n}  \frac{\partial^n}{\partial( \lambda^*)^n} \left \langle{ 0_q \left| {\exp \left[{-\lambda^* b_q }\right]  \exp \left[{ \lambda b_q^{\dagger} }\right] }\right| 0_q }\right  \rangle =P_n \left( { \lambda, \lambda^*}\right) \exp\left[ {- |\lambda|^2 }\right] \;\; ,
\]

\noindent with $P_n$ a polynomial expression, to the result that 

\begin{mathletters}
\label{eq28}

\begin{equation}
\langle \{ n_{q, J} \} | \{ n_{q, J'} \} \rangle \propto \langle 0_J| 0_{J'} \rangle =\exp\left[ - \frac{1}{2} \sum\limits_{q} {\left|{ \frac{\mu_{Jq}-\mu_{J'q}}{\hbar \omega_q} }\right |^2 } \right] 
\end{equation}

\noindent and, equivalently,

\begin{equation}                                                                                                                                                                      
\exp \left[ { -\Gamma_{J, J'}^0 (T) } \right] \propto  \langle 0_J | 0_{J'} \rangle  = \exp \left[ { -\Gamma_{J,J'}^0 (T=0) } \right]\;.
\end{equation}

\end{mathletters}

\noindent As in Sec.IIIB, here $ | 0_J \rangle$ denotes the displaced environment vacuum created by the $J$-th DFS. 

Similarly to the pure state case therein, the finite character of the stationary dissipative factor $\Gamma_{J, J'}^0 (T)$ is seen to be conditioned by the {\it unitarily equivalent or inequivalent} character of the environment vacua generated by the register DFS, or equivalently, by its zero temperature value $\Gamma_{J,J'}^0 (T=0)$ [this latter form of the condition can be obtained directly from expression (\ref{eq26}) for  $\Gamma_{J, J'}^0 (T)$ in the zero temperature limit]. Clearly, the displaced vacua are unitarily inequivalent [have vanishing overlap], and $\Gamma_{J, J'}^0 (T=0)$ also vanishes, when  $\langle 0_{q,J} |0_{q,J'} \rangle \propto \exp[-(1/2) |\mu_{Jq} - \mu_{J'q} |^2/(\hbar\omega_q)^2] \to 0$ as $q \to 0$ [$|(\mu_{Jq} - \mu_{J'q} )/\hbar\omega_q|$ diverges as $q \to 0$], unless the density of modes compensates for the contributions from low-frequency modes. At finite temperatures the effect is further amplified by the thermal excitation of low-frequency states. On the other hand, the situation can improve considerably when $|(\mu_{Jq} - \mu_{J'q} )/\hbar\omega_q|$ remains finite as $q \to 0$, and the displaced vacua remain in unitarily equivalent Fock spaces. \\

We may conclude that decoherence free states involving multiple entangled DFSs are possible in the thermal coherent-product ansatz if conditions (i)-(ii) above are simultaneously satisfied alongside condition \\

\noindent iii) {\it the stationary part $\Gamma_{J, J'}^0 $ of the dissipative factors $\Gamma_{J\nu, J'\nu'}$ must be finite; a necessary prequisite is that the environment displaced vacua associated with the contributing DFSs belong to unitarily equivalent Fock spaces.} \\

\noindent Conditions (i) and (iii) impose limiting restrictions on the register-environment coupling, by demanding both a particular structure of the coupling constants, and a suitable density of environment states. Condition (ii) can be always satisfied for continuously degenerate environment modes, by a proper choice of initial displacements. Finite degenerate modes [e.g., the 2-fold degenerate modes of a 1-dimensional environment] can also comply with condition (ii), provided the number of distinct displacements is appropriately limited.  

\section{Examples: the spin-boson model}

Let us illustrate the arguments of Sec.IV on the specific case of model (\ref{eq1}). Under a common functional prescription for the density of states, it turns out that a single qubit register [$N=1$, $\chi_{qn} = \chi(\omega_q)$] can display a finite stationary dissipative factor $\Gamma^0$ only in a 3-dimensional environment. Not surprisingly, a similar behavior is also seen in multiqubit registers with weak collective decoherence [$N>1$, $\chi_{qn} = \chi(\omega_q)$]. However, linear registers with individual decoherence [$\chi_{qn}=\chi(\omega_q) \exp (iq \cdot r_n)$] present states for which $\exp[-(1/2) |\mu_{Jq} - \mu_{J'q} |^2/(\hbar\omega_q)^2]$ remains finite as  $q \to 0$, and which generate finite $\Gamma^0$-s even in a 1-dimensional environment.

\subsection{Single-qubit register}

For a single qubit register [$N=n=1$] let the model coupling constants assume the form

\[
\chi_{q} = \chi(\omega_q)\;.
\]

\noindent There are only two one-dimensional DFS, such that $J\nu=\uparrow$ or $J\nu=\downarrow$, and the associated couplings $\mu_{Jq}$ become $\mu_{\uparrow,q}=-\mu_{\downarrow,q}=\chi(\omega_q)$. Assuming a total coherent-product state, the reduced register state (\ref{eq16}) has the form

\[
\hat\rho^R(t) = e^{-\frac{i}{\hbar}\varepsilon \sigma_z t} \cdot \left[ { |\uparrow\rangle \langle \uparrow | \hat\rho^R| \uparrow \rangle \langle \uparrow| + |\downarrow\rangle \langle \downarrow | \hat\rho^R| \downarrow \rangle \langle \downarrow| + }\right.
\]

\begin{equation}
\label{eq28b}
\left.{ + \eta(t) |\uparrow\rangle \langle \uparrow | \hat\rho^R| \downarrow \rangle \langle \downarrow| + \eta^* (t) |\downarrow\rangle \langle \uparrow | \hat\rho^R| \uparrow \rangle  \langle \uparrow | \hat\rho^R| \downarrow \rangle^* \langle \uparrow| }\right] \cdot e^{\frac{i}{\hbar}\varepsilon \sigma_z t} \;\;,
\end{equation}

\noindent with the environment driven correlation $\eta(t)$ is given by Eq.(\ref{eq18a}) with appropriate substitutions. 

The state $\hat\rho^R(t)$ becomes decoherence free provided $\eta(t) = const.$ under the corresponding conditions (i)-(iii) of Sec.IV. Condition (i) [Eq.(\ref{eq23}a)] is satisfied by default, due to the particular form of the coupling constants. Condition (ii) [Eqs.(\ref{eq24})] reduces to 

\begin{mathletters}
\label{eq29}

\begin{equation}
\oint \limits_{S_\omega} {dS_q \beta_{\uparrow q}^0} = \oint \limits_{S_\omega} {dS_q \beta_{\downarrow q}^0} =0\;,
\end{equation}

\noindent while condition (iii) for a finite stationary part of the dissipative factor $\Gamma$ requires [see Eq.(\ref{eq26})] 

\begin{equation}
\Gamma^0 \equiv \frac{1}{2} \int{d\omega \frac{d|q|}{d\omega} (\hbar \omega)^{-2} \coth \left( {\frac{\hbar \omega }{2k_B T}} \right) \oint \limits_{S_\omega} {dS_q  \left|{ 2 \chi(\omega)} \right|^2  } } = \int{d\omega \frac{d|q|}{d\omega}G(\omega) \left| {\frac{ \chi(\omega) }{\hbar \omega} }\right|^2 \coth \left( {\frac{\hbar \omega }{2k_B T}} \right)  } < \infty \;,
\end{equation}

\end{mathletters}

\noindent where $G(\omega)$ is the density of modes at frequency $\omega$. 

The density $G(\omega)$ grows as $area(S_\omega) = \oint \limits_{S_\omega} dS_q \sim \omega^{d-1}$, where $d$ is the dimension of the environment, and is characterized by a natural ultraviolet cut-off frequency $\omega_c$, which sets the upper limit for the rate of dissipation processes in the environment. This necessary feature is usually accounted for by setting $G(\omega) \propto \omega^{d-1} \exp[-\omega/\omega_c]$. It is also common to assume a quasi-linear dispersion $d|q|/d\omega \sim const.$ and a coupling $\chi(\omega) \propto \sqrt \omega$. Alternatively, one can transfer the cut-off exponential onto the interaction strength, and set $|\chi(\omega)|^2 \propto \omega \exp[-\omega/\omega_c]$, while leaving the density of states as $ G(\omega) \sim \omega^{d-1}$. In either case the final prescription reads

\[
(d|q|/d\omega)G(\omega) |\chi(\omega)|^2 =\lambda (\hbar^2 \omega_c/2) (\omega/\omega_c)^d exp[-\omega/\omega_c] \;\;,
\]

\noindent where the scaling factors are chosen such that $\lambda$ is an adimensional constant. In the following we favor the second interpretation, which can be straightforwardly extended to cases with couplings unisotropic in the wavevector $q$.

For the present single-qubit case, $|\chi(\omega)/\hbar\omega|^2 \propto \omega^{-1} \exp[-\omega/\omega_c] \to \infty$ as $q \to 0$, and the stationary dissipative factor $\Gamma^0$ cannot be finite [equivalently, the displaced environment vacua associated with each of the two qubit states cannot be unitarily equivalent] unless this singularity is balanced by the density of states. This is also evident from the integrand in Eq. (\ref{eq29}b), which behaves in the low-frequency limit as $\omega^{d-3}$. In particular, a 1-dimensional density of states proves insufficient to counteract the low-frequency contribution even at zero temperature, i.e., the displaced environment vacua are unitarily inequivalent. A 2-dimensional density suffices at zero temperature, implying that the displaced environment vacua are unitarily equivalent, but fails at finite temperatures. Only in a 3-dimensional environment does $\Gamma^0$ acquire finite values at both zero and finite temperatures, meaning that the displaced vacua are unitarily equivalent and the generated DFS correlations are stable under thermal excitation. In this case, the exact expression of $\Gamma^0$ can be given an analytical form in terms of the generalized Riemann zeta function $\zeta(g, z)=(1/ \tilde\Gamma (g) ) \int \limits_{0}^{\infty} {d\xi \xi^{g-1} e^{-z \xi} /(1-e^{-\xi}) }$ [here $ \tilde\Gamma (g)$ denotes the Gamma function] as 

\begin{equation}
\label{eq30}
\Gamma^0 \left( { \frac{k_B T}{\hbar \omega_c} }\right) = \Gamma^0 (0) \left[ { 2 \left( { \frac{k_B T}{\hbar \omega_c} }\right)^2 \zeta \left( 2, { \frac{k_B T}{\hbar \omega_c} }\right) -1 } \right] \;,
\end{equation}

\noindent where $\Gamma^0 (0) = \lambda$. 

Expression (\ref{eq30}) shows that the stationary factor $\Gamma^0$ increases monotonously with the temperature, behaving as $\Gamma^0 \approx \Gamma^0(0) \left[ { 1+ \left( {\pi^2/3} \right) \left( { k_B T/ \hbar \omega_c} \right)^2 } \right]$ in the low temperature limit [$k_B T << \hbar\omega_c$] and as  
$\Gamma^0 \approx 2 \Gamma^0(0) \left( {k_B T/ \hbar \omega_c} \right)$ at high temperatures [$k_B T >> \hbar\omega_c$]. Accordingly, the maximum amplitude of register correlations $( \eta \cdot \langle \uparrow | \hat\rho^R| \downarrow \rangle)$ that can be carried in an environment-entangled unperturbed propagation decreases fast to zero in a {\it classical} environment, with temperatures $k_B T \geq \hbar\omega_c$. A nontrivial unperturbed propagation is seen to require a low-temperature, {\it quantum} environment. The widest accessible range of register correlations is supplied therefore by zero-temperature, Davydov-like pure states of the joint register-environment system. For environment displacements $\beta_{\uparrow (\downarrow) q}(t)$ satisfying the phasing condition (\ref{eq29}a), it can be checked that such states acquire the form

\[
|\Phi(t)\rangle = e^{-\frac{i}{\hbar}\varepsilon \sigma_z t} e^{\frac{i}{\hbar}\Omega t} \left[{ c_{\uparrow} |\uparrow\rangle \otimes |\beta_{\uparrow}(t)\rangle + c_{\downarrow} |\downarrow\rangle \otimes |\beta_{\downarrow}(t)\rangle}\right]\;\;,
\]

\noindent with $ |\beta_{\uparrow (\downarrow)}(t)\rangle = exp\left[{ \sum\limits_q { \left({\beta_{\uparrow (\downarrow) q}(t)b^{\dagger}_q -\beta_{\uparrow (\downarrow) q}(t)^* b_q }\right) } }\right] |0_B\rangle$, and $\langle \beta_{\uparrow}(t)|\beta_{\downarrow}(t)\rangle = \langle \beta_{\uparrow}(0) | \beta_{\downarrow} (0)\rangle $. It was also taken into account that under condition (\ref{eq29}a) the phase factors $\Theta_{\uparrow (\downarrow)}(t)$ reduce to $\Theta_{\uparrow}(t) = \Theta_{\downarrow}(t) = \frac{1}{\hbar}\Omega t$, with $\Omega = -\int d\omega (d|q|/d\omega) G(\omega) [|\chi(\omega)|^2/\hbar\omega ] $.

\subsection{N-qubit register with weak collective decoherence}

The situation is very similar in the closely related case of a multi-qubit register [$N>1, 0 \leq n \leq (N-1) $] with weak collective decoherence, when again 

\[
\chi_{qn}=\chi(\omega_q)\;\;.
\]

\noindent The register DFSs are now indexed by the total spin projection on the z-axis, e.g.,

\[
J = \sum \limits_{n=0}^{N-1}{s^{(n)}_\nu} \;\;,
\]

\noindent where $s^{(n)}_\nu$ is the eigenvalue of $\sigma_z^{(n)}$ labeling the $\nu$-th state of the $J$-th DFS $\cal S_J$, and the corresponding coupling constants read 

\[
\mu_{Jq} = J\cdot\chi(\omega_q)  \;\;.
\]

\noindent Due to condition (\ref{eq23}a), the register states contributing to a given environment-entangled decoherence-free distribution can only belong to either of a pair of DFS, $\{ {\cal S}_J \cup {\cal S}_{-J} \}$, characterized by coupling constants $J\cdot\chi(\omega_q)$ and $(-J)\cdot\chi(\omega_q)$, respectively. As a result, an entangled decoherence free state is of the form

\[
\hat\rho^R(t) = \exp\left[{ -\frac{i}{\hbar}H_Rt }\right] \cdot \sum\limits_{\nu, \nu'} { \left[{ \;\eta_{J\nu,J\nu'} P_{J\nu}\hat\rho^R P_{J\nu'} +\eta_{-J\nu,-J\nu'} P_{-J\nu}\hat\rho^R P_{-J\nu'}+ }\right.}
\]

\begin{equation}
\label{eq30b}
\left.{ +  \eta_{J\nu,-J\nu'}(t) P_{J\nu}\hat\rho^R P_{-J\nu'}+ \eta_{J\nu,-J\nu'}^*(t) P_{-J\nu}\hat\rho^R P_{J\nu'}\; }\right]  \cdot \exp\left[{ \frac{i}{\hbar}H_Rt }\right]\;\;,
\end{equation} 
$ \;$

\noindent where the bath correlations between states in the same DFS are explicitly shown to be time-independent. The cross-DFS bath correlations $\eta_{J\nu,-J\nu'}(t)$ become time independent provided  the environment displacements are such that [see Eqs.(\ref{eq24})]

\begin{equation}
\label{eq31b}
\oint \limits_{S_\omega} {dS_q \beta_{J\nu,q}^0} = \oint \limits_{S_\omega} {dS_q \beta_{-J\nu,q}^0} =0\;, \;\forall \nu\;\;,
\end{equation}

\noindent and remain nonvanishing if the stationary dissipative factors $\Gamma^0_{J,-J}$ are finite, which amounts to  

\begin{equation}
\label{eq32b}
\Gamma^0_{J,-J} \equiv \frac{1}{2} \int{d\omega \frac{d|q|}{d\omega} (\hbar \omega)^{-2} \coth \left( {\frac{\hbar \omega }{2k_B T}} \right) \oint \limits_{S_\omega} {dS_q  \left|{ 2 J \chi(\omega)} \right|^2  } } = J^2 \Gamma^0 < \infty \;.
\end{equation}

\noindent In Eq.(\ref{eq32b}) above, $\Gamma^0$ is the stationary dissipative factor discussed previously for a single qubit. Evidently, under the same prescription for the density of environmental modes, a similar discussion applies and shows that the bath correlations $\eta_{J\nu,-J\nu'}$ can be constant and nonzero only in a 3-dimensional environment. Note, however, that their magnitude decreases strongly for register states with spin projections $ \left| J \right| > 1$. In the zero-temperature, pure state limit, decoherence free register distributions are generated by Davydov-like states of the total register-environment system with at most ${\it dim} ({\cal S}_J) + {\it dim} ({\cal S}_{-J}) = 2 \times {\it dim} ({\cal S}_J)$ distinct terms.

\subsection{N-qubit register with individual decoherence}

Consider now a N-qubit register [$N > 1, 0\leq n\leq(N-1)$] in an individual decoherence regime, characterized by coupling parameters 

\[
\chi_{qn}=\chi(\omega_q) \exp (iq \cdot r_n) \;\;,
\]

\noindent where $r_n$ denotes the position vector of the $n$-th qubit. We also assume, for convenience, a linear geometry, such that 

\[
r_n=n a \;\;.
\]

\noindent The DFS of this register are trivially 1-dimensional and correspond to the eigenstates $| \{s^{(n)}_J \} \rangle$ of the unperturbed $H_R$, such that the coupling constants $\mu_{Jq}$ now amount to

\[
\mu_{Jq} = \chi(\omega_q) \sum \limits_{n=0}^{N-1}{ e^{in q a}s^{(n)}_J}\;\;.
\]

\noindent Note that the DFS index $J$ is no longer associated to the total spin projection as in the previous case, but labels individual unperturbed eigenstates. 

Select a specific eigenstate $| \{s^{(n)}_{J_0} \} \rangle$. In order to detect a set of entangled decoherence free states involving $| \{s^{(n)}_{J_0} \} \rangle$, consider first the corresponding condition (\ref{eq23}a), which limits the set of compatible coupling constants $\mu_{Jq}$. A straightforward solution is to seek states for which $|\mu_{Jq}| = |\mu_{J_0 q}|, \forall q$. Given the linear translational symmetry of the register, it is not hard to observe that the set of such states includes the $N$ states $| \{s^{(n)}_{J_m} \} \rangle$ [$0 \geq m \geq (N-1)$] related to $| \{s^{(n)}_{J_0} \} \rangle$ by a cyclic permutation modulo N, i.e. 

\[
s^{(n)}_{J_m}=s^{(n-m)}_{J_0}\;\;, \; \forall n\;\;.
\]

\noindent Indeed, the corresponding coupling constants differ from $\mu_{J_0 q}$ only by a phase factor, respectively

\[
\mu_{J_m q}=e^{iqma} \mu_{J_0 q}\;\;.
\]

\noindent A coherent-product total state involving the $N$ register states thus selected produces a reduced register state [see again Eq.(\ref{eq16})]

\begin{equation}
\label{eq33b}
\hat\rho^R(t) = \exp\left[{ -\frac{i}{\hbar}H_Rt }\right] \cdot \sum\limits_{m, m'} { \eta_{J_m,J_{m'}}(t) P_{J_m}\hat\rho^R P_{J_{m'} } } \cdot \exp\left[{ \frac{i}{\hbar}H_Rt }\right]\;\;,
\end{equation} 

\noindent for which the environment mediated correlations $\eta_{J_m,J_{m'}}(t)$ become constant in time when the bath displacements comply with [Eq.(\ref{eq24})]

\begin{equation}
\label{eq34b}
\oint\limits_{S_\omega  } {dS_q \beta_{J_m, q}^0 \left( {\mu_{J_m q}^*  - \mu_{J_{m'}q}^* } \right) }  = 0\;, \;\forall m, m'\;\;.
\end{equation}

\noindent Nonvanishing $\eta_{J_m,J_{m'}}$-s require in addition a finite corresponding dissipative factor, i.e. [Eq.(\ref{eq26})]

\begin{equation}
\label{eq35b}
\Gamma_{J_m, J_{m'}}^0  \propto  \frac{1}{2} \int{d\omega \frac{d|q|}{d\omega} \coth \left( {\frac{\hbar \omega }{2k_B T}} \right) \oint \limits_{S_\omega} {dS_q  \left| { \frac{\mu_{J_m q}  - \mu_{J_{m'} q} }{\hbar\omega} } \right|^2  } } < \infty
\end{equation}

\noindent Let us assume again the usual prescription for the density states, as described for the single-qubit case. If we also introduce the transit time $t_s$, defined by $qa=\omega t_s$, it can be seen that for any $m$ and $m'$ the integrand under the surface integral above reads

\[
|(\mu_{J_m q} - \mu_{J_{m'} q} )/\hbar\omega_q|^2 = 4 |(\chi(\omega)/\hbar\omega) \sin[(m'-m)\omega t_s /2] \sum\limits_{n=0}^{N-1}{e^{i n \omega t_s}s^{(n)}_{J_0} } |^2\;\;,
\]

\noindent and behaves as $\omega $ when $q \to 0$ [recall that $|\chi(\omega)|^2 \sim \omega \exp[-\omega/\omega_c]$ ]. Hence the entire integrand in the stationary dissipative factor $\Gamma_{J_m, J_{m'}}^0$ behaves in the low-frequency limit as ${\it area}(S_\omega) \sim \omega^{d-1}$ [for $d|q|/d\omega \sim const.$] and $\Gamma _{ J_m, J_{m'}}^0$ remains finite in any environment, at any finite temperature. Equivalently, the displaced environment vacua corresponding to the register states $| \{s^{(n)}_{J_m} \} \rangle$ and $| \{s^{(n)}_{J_{m'}} \} \rangle$ are unitarily equivalent and the corresponding correlations between the register states are thermally stable.

One can also consider the mirror permutation states with 

\[
s^{(n)}_{\overline J_m}=s^{(N-n)}_{J_m}=s^{(N-n+m)}_{J_0}\;\;,
\]

\noindent and 

\[
\mu_{\overline J_m q}=e^{i(N-m)qa} \chi(\omega_q)\left({\frac{\mu_{J_0 q}}{\chi(\omega_q)} }\right)^*\;\;,
\]

\noindent which are compatible with condition (\ref{eq23}a) as well. The equilibrium dissipative factor of any pair of such states reads  $\Gamma _{ \overline J_m, \overline J_{m'}}^0=\Gamma _{J_m, J_{m'}}^0$, while the factor corresponding to a direct permutation state and a mirror permutation state is given by

\begin{equation}
\label{eq32}
\Gamma _{J_m \overline J_{m'} }^0  \propto 2\int\limits_0^\infty  {d\omega \frac{d\left| q \right|}{d\omega } \coth \left( {\frac{\hbar \omega }{2k_B T}} \right)} {\rm  }\oint\limits_{S_\omega  } {dS_q \left| {\frac{\chi (\omega )}{\hbar \omega }\sum\limits_{n = 0}^{N - 1} {s^{(n)}_{J_0} \sin \left[{ \left({ n+\frac{m+m'-N}{2} }\right) \omega t_s }\right] } } \right|^2 } \;,
\end{equation} 

\noindent and again proves to be finite in any environment. It follows that any environment-entangled distributions which involve the $2N$ direct and mirror permutation counterparts of  $| \{s^{(n)}_{J_0} \} \rangle$ and satisfy conditions (\ref{eq34b}) generate nontrivial register mixtures propagating in an unperturbed manner. Note however that in a 1-dimensional environment such a mixture can only accommodate two distinct pure states [see Eqs. (\ref{eq34b})]. It is also worth noting that in this case the precise magnitude of the stationary dissipative factor $\Gamma^0$ is strongly dependent on the input states. The simplest example of a joint register-environment state yielding a decoherence free register evolution is again a Davydov-like pure state, which in the present case takes the form

\[
|\Phi(t) \rangle = \exp\left[{ - \frac{i}{\hbar}} \sum\limits_n {\varepsilon \sigma_z^{(n)}} t\right] e^{i\Omega_{J_0}t} \sum\limits_m {\left[{c_m |s_{J_m}^{(n)}\rangle \otimes |\beta_{J_m} (t) \rangle + \overline c_m |s_{\overline J_m}^{(n)}\rangle \otimes | \beta_{\overline J_m} (t)}\rangle\right]}\;\;.
\]

\noindent The expression above is obtained for coherent environment states $|\beta_{J_m} (t) \rangle$ and $|\beta_{\overline J_m} (t)\rangle$ with displacements $\beta_{J_m, q} (t)$, $\beta_{\overline J_m, q} (t)$ satisfying synchronization conditions similar to Eq.(\ref{eq34b}). In this case the overlaps $\langle \beta_{J_m} (t)| \beta_{J_{m'}} (t)\rangle$,  $\langle \beta_{\overline J_m} (t)| \beta_{\overline J_{m'}} (t)\rangle$ and $\langle \beta_{J_m} (t)| \beta_{\overline J_{m'}} (t)\rangle$ are constant in time and the phase factors $\Theta_{J\nu}(t)$ [see Eq.(\ref{eq14}a)] become $\Theta_{J_m}(t) = \Theta_{\overline J_m}(t)=\Theta_{J_0}(t) = (1/\hbar) \Omega_{J_0} t, \forall m$, with $\Omega_{J_0}$ a constant energy shift appropriately defined.\\ \\ \\

\section{Elementary entanglement rephasing through qubit flipping: \\ continuos-time coherence evolution during bang-bang control}

The coherent-product ansatz underlying our discussion of multi-DFS decoherence free states was introduced in Sec.III via a {\it formal} preparation protocol relying on an explicit alteration of the register interaction with the environment. But a {\it practical} approach to the preparation problem may not necessarily require direct manipulation of the environment. As a counterexample, let us point out that at least one method of decoherence control available in the literature involves an elementary rephasing of entanglement of the type discussed here. We refer specifically to quantum bang-bang control \cite{QBB}, inspired by the multi-pulse decoupling techniques of NMR, which counteracts decoherence through a train of identical spin-flip cycles. Each cycle generates a revival of coherence through a pair of coherent $\pi$-pulses that alternately flip the state of the register. When this process is examined in the context of the extended density matrix solution of Sec. III, the cause of this revival effect is distinctly exposed as an elementary adjustment of the entanglement with the environment. In other words, spin flipping provides a simple working procedure for manipulating entanglement.  In a supplementary outcome, the exact density matrix solution brings forth an alternative picture on the source of the refining power of so-called symmetric cycles \cite{Viola}. The latter introduce a straightforward adjustment of the bang-bang technique, which lowers the working cycle frequency by a factor of at least 2.   

For simplicity, consider only model (\ref{eq1}) with a single qubit as described in Sec.IVA. A bang-bang procedure applies a succession of resonant radiofrequency pulses [$\pi$-pulses] which evolve one eigenstate of the qubit into the other, i.e., $|\uparrow \rangle \to |\downarrow \rangle$ and $| \downarrow \rangle \to |\uparrow \rangle$, on a time scale $\tau_P$ short compared to the bath correlation time. As before, throughout the following we assume the Schroedinger picture. If the rf field is strong enough, the interaction of the qubit with the environment can be neglected during the pulse, so that the bath coordinates remain unaffected. Therefore a total environment-entangled state, which reads [Eq.(\ref{eq12}) with a convenient rearrangement]  

\[ 
\hat \rho (t_ -  ) = \left|  \uparrow  \right\rangle \hat \rho _{e, \uparrow  \uparrow } \left\langle  \uparrow  \right| + \left|  \downarrow  \right\rangle \hat \rho _{e, \downarrow  \downarrow } \left\langle  \downarrow  \right| + \left|  \uparrow  \right\rangle \hat \rho _{e, \uparrow  \downarrow } \left\langle  \downarrow  \right| + \left|  \downarrow  \right\rangle \left( {\hat \rho _{e, \uparrow  \downarrow } } \right)^{\dagger}  \left\langle  \uparrow  \right|
\]

\noindent immediately before the pulse, transforms into 

\[
\hat \rho (t_ + ) = \left|  \downarrow  \right\rangle \hat \rho _{e, \uparrow  \uparrow } \left\langle  \downarrow  \right| + \left|  \uparrow  \right\rangle \hat \rho _{e, \downarrow  \downarrow } \left\langle  \uparrow  \right| + \left|  \downarrow  \right\rangle \hat \rho _{e, \uparrow  \downarrow } \left\langle  \uparrow  \right| + \left|  \uparrow  \right\rangle \left( {\hat \rho _{e, \uparrow  \downarrow } } \right)^{\dagger}  \left\langle  \downarrow  \right| 
\]

\noindent immediately after the pulse. An elementary spin-flip cycle consists of two such pulses applied at a time interval $\Delta t$, the first of which reverses the qubit state, while the second restores the original configuration. Between any two pulses the qubit-environment system resumes the dynamics described by Hamiltonian (\ref{eq1}). For the purpose of illustration, it is sufficient to examine the process in the limit of infinitely narrow pulses, $\tau_P \to 0$ \cite{QBB}, when each cycle can be approximated by the following piecewise evolution: i) propagation under Hamiltonian (\ref{eq1}) for a duration $\Delta t$, starting at time $t_n$; ii) instantaneous $\pi$-pulse and interchange of qubit eigenstates states at time $(t_n + \Delta t)$; iii) evolution under Hamiltonian (\ref{eq1}) from time $(t_n + \Delta t)$ to time $t_{n+1}=t_n +2 \Delta t$; iv) second $\pi$-pulse and interchange of qubit eigenstates at time $t_{n+1}$. 

The analysis of this process in the framework of solution (\ref{eq3}) is quite straightforward provided we shift focus from the qubit eigenstates to the entangled environment distributions. Indeed, the transformation of a qubit-environment state upon application of a $\pi$-pulse can be interpreted also as a switch of the environment distributions associated with the qubit states, in the sense that $\hat \rho _{e, \uparrow  \uparrow }  \to \hat \rho _{e, \downarrow  \downarrow } $, $\hat \rho _{e, \uparrow  \downarrow }  \to \hat \rho _{e, \uparrow  \downarrow } $, etc. The latter implies that the associated bath displacements before the rf pulse for $\hat\rho_{e,\uparrow \uparrow}$, etc., become initial displacements for $\hat\rho_{e,\downarrow \downarrow}$, etc., after the pulse and vice versa, i.e.,

\[
\beta _{\uparrow q } (n \Delta t+0 ) = \beta _{\downarrow q} (n \Delta t -0 )\;\;,
\]

\noindent and 

\[
\beta _{\downarrow q} (n \Delta t +0 ) = \beta _{\uparrow q} (n \Delta t -0 )\;\;.
\]

\noindent For this reason, the interchange of qubit eigenstates following a $\pi$-pulse is equivalent to a rephasing of the entangled environment distributions. 

The time-dependence of the bath displacements for the entangled $\hat \rho_e$-s ensues now without difficulty. Starting at time $t_n=2n\Delta t$ [$t_0 =0$] with displacements $\beta_{\uparrow (\downarrow) q}(t_n)$, one has [see Eq.(\ref{eq15})]:

\noindent i)  for $t_n< t < (t_n+\Delta t)=(2n+1)\Delta t$: $\beta _{\uparrow (\downarrow) q} (t) = \left[ {\beta _{ \downarrow (\uparrow) q} (t_n ) \pm \chi _q /\hbar \omega _q } \right]\exp [ - i\omega _q (t - t_n )] \mp \chi _q /\hbar \omega _q $; 

\noindent ii) for $(2n+1)\Delta t < t < t_{n+1}=2(n+1)\Delta t$: $\beta _{\uparrow (\downarrow) q} (t) = \left[ {\beta _{\downarrow (\uparrow) q} (t_n +\Delta t) \pm \chi _q /\hbar \omega _q } \right]\exp [ - i\omega _q (t - t_n-\Delta t )] \mp \chi _q /\hbar \omega _q $.

\noindent Further, let the initial qubit-environment state be unentangled, with the environment in thermal equilibrium, such that $\beta _{\uparrow q} (0)=\beta _{\downarrow q} (0)=0$. Since the unperturbed evolution in the first half-cycle drives same mode displacements to values of opposite sign, such that  $\beta _{\uparrow q} (t)=-\beta _{\downarrow q} (t)=(\chi_q/\hbar\omega_q) [ \exp(-i \omega_q t) - 1]$ for $ 0<t<\Delta t $, the corresponding evolved displacements will have opposite signs at any later time, that is, $\beta _{\uparrow q} (t)=-\beta _{\downarrow q}(t)$ for all $t$. As a result, all phase factors [Eqs.(\ref{eq14}a) and (\ref{eq19a}a)] in the bath correlation factor $\eta(t)$ for the reduced qubit state (\ref{eq28b}) vanish, and the only relevant quantity remains the dissipative factor [Eq.(\ref{eq19a}b)]

\begin{equation}
\label{eq33}
\Gamma(t)=2\sum\limits_q {|\beta_{\uparrow q}(t)|^2 \coth(\hbar\omega_q/2k_b T)}\;\;.
\end{equation}

\noindent At the same time, the effect of a $\pi$-pulse is seen to amount to a change of sign of the bath displacements in the entangled environment distributions, such that 

\[
\beta _{\uparrow (\downarrow) q} (k\Delta t +0)=-\beta _{\uparrow (\downarrow)q } (k\Delta t -0)\;\;.
\]

\noindent Accounting for this into the time-dependence of the displacements, yields immediately stroboscopic recurrence relations of the form 

\begin{equation}
\label{eq34}
\beta _{\uparrow q} (k \Delta t+0) = \beta _{\uparrow q} ((k-2) \Delta t +0)e^{ - 2i\omega _q \Delta t}  + \frac{{\chi _q }}{{\hbar \omega _q }}\left( {e^{ - i\omega _q \Delta t}  - 1} \right)^2 \;.
\end{equation}

\noindent Solving the recurrence for $k=2n$ with initial condition $\beta _{\uparrow q} (0)=0$ recovers the stroboscopic result for $\Gamma(t)$ obtained in ref.\cite{QBB}, which we write in the form

\[
\Gamma_{strob} (2n\Delta t)=2\sum \limits_q { \coth\left({ \frac{\hbar\omega_q}{2 k_B T} }\right) \left| {\beta _{\uparrow q,0}(2n\Delta t)\; \tan \left({ \frac{\omega_q \Delta t}{2} }\right) }\right|^2 } \;,
\] 

\noindent where $\beta _{\uparrow q, 0} (t)=(\chi_q/\hbar\omega_q) [ \exp(-i \omega_q t) - 1]$ denotes the evolved displacement in the absence of pulses. Under the usual prescription for the density of states [see previous section], the above expression becomes 

\begin{equation}\label{eq35}
\Gamma_{strob} (2n\Delta t)=\lambda \hbar^2 \omega_c \int\limits_{0}^{\infty} { d\omega \left( \frac{\omega}{\omega_c} \right)^{d-1} \exp\left({-\frac{\omega}{\omega_c} }\right) \coth\left({ \frac{\hbar\omega}{2k_B T} }\right) \left | {\beta _{\uparrow q,0}(2n\Delta t)\; \tan \left({ \frac{\omega_q \Delta t}{2} }\right) }\right|^2 } \;.
\end{equation} 

\noindent The decay of the corresponding bath correlation factor [usually referred to as qubit coherence] $\eta_{strob} (2n\Delta t) = \exp[-\Gamma_{strob} (2n\Delta t)]$ was shown to be strongly suppressed at high enough pulse frequencies [$\omega_c \Delta t \leq 1$], and to become completely quenched in the limit of continuous flipping. The effect is interpreted in ref.\cite{QBB} as an approximate time reversal of the qubit evolution induced by each $\pi$-pulse, which becomes visible as an effective change of sign of the total Hamiltonian. The present point of view adds that the rephasing of bath entanglement responsible for the partial time reversal occurs precisely through a change of sign of the bath displacements and so amounts to a change of sign of the time derivative of the dissipative factor, since

\begin{equation}
\label{eq36}
\frac{d \Gamma }{dt} = 2i \sum\limits_q {\omega _q \coth \left( {\frac{\hbar \omega _q }{2k_B T} } \right)\left[ {\frac{\chi _q^* }{\hbar \omega _q }\beta _{\uparrow q} (t) - \frac{\chi _q }{\hbar \omega _q }\beta _{\uparrow q}^* (t)} \right]} \;.
\end{equation}
    
\noindent In contrast, the exact time reversal of $\Gamma(t)$ at time $t$ requires that the bath displacements change according to $\chi_q^* \beta_{\uparrow (\downarrow) q}(t +)=\chi_q [\beta_{\uparrow (\downarrow) q}(t -)]^*$. Remarkably, this differs from the flip-induced change by a mere phase factor [see also \cite{QBB}]. Thus if the bath correlation factor $\eta=\exp(-\Gamma)$ decays during the last moments of the first half-cycle, a change of sign of the displacements after the mid-cycle flip suffices to induce a subsequent revival of coherence. Formal evidence for this phenomenon is provided by the expression of the time-derivative $d\Gamma/dt$ at the moment immediately following the mid-cycle pulse, which can be shown straightforwardly to amount to

\begin{equation}
\label{eq37}
\frac{d\Gamma}{dt}((2n-1)\Delta t+0) = -\lambda \hbar^2 \omega_c \int\limits_{0}^{\infty} { d\omega \left( \frac{\omega}{\omega_c} \right)^{d-1} \exp\left({-\frac{\omega}{\omega_c} }\right) \coth\left({ \frac{\hbar\omega}{2k_B T} }\right) \cos^2\left({ (n-\frac{1}{2})\omega \Delta t }\right) \tan \left({\frac{\omega\Delta t}{2}}\right)}
\end{equation}

\noindent for a standard density of states. It is not difficult to observe that for $d \leq 3$ this expression is always negative, since the dominant contribution comes from the range $0\leq \omega \Delta t \leq \pi$, where the integrand is negative [note that for $\omega \Delta t \geq \pi$ the periodical factor $\cos^2((n-\frac{1}{2})\omega \Delta t) \tan \left({\frac{\omega\Delta t}{2}}\right)$ is multiplied by a monotonously decreasing function if $d \leq 4$]. Hence the correlation factor $\eta$ always increases after the mid-cycle flip. 

Surprisingly, the detailed solution reveals that the second flip also induces a revival. Indeed, the derivative of $\Gamma$ at the moment immediately preceding the second pulse of a cycle reads

\begin{equation}
\label{eq38}
\frac{d\Gamma}{dt}(2n\Delta t-0) = \lambda \hbar^2 \omega_c \int\limits_{0}^{\infty} { d\omega \left( \frac{\omega}{\omega_c} \right)^{d-1} \exp\left({-\frac{\omega}{\omega_c} }\right) \coth\left({ \frac{\hbar\omega}{2k_B T} }\right) \sin^2(n\omega \Delta t) \tan \left({\frac{\omega\Delta t}{2}}\right)}
\end{equation}
 
\noindent  and, by the same argument as for expression (\ref{eq37}), is noted to be always positive for $d \leq 3$, since the integrant is positive in the dominant range $0\leq \omega \Delta t \leq \pi$. Because $\eta(t)$ is necessarily decreasing at this moment, the second pulse can only produce a revival. It also becomes apparent that $\eta$ displays [at least] a maximum between any two consecutive pulses.

This effect is confirmed by a numerical integration of the exact dissipative factor (\ref{eq33}), which becomes, for the selected density of states,

\begin{equation}
\label{eq38b}
\Gamma(t)=\lambda \hbar^2 \omega_c \int\limits_{0}^{\infty} { d\omega \left( \frac{\omega}{\omega_c} \right)^{d-1} \exp\left({-\frac{\omega}{\omega_c} }\right) \coth\left({ \frac{\hbar\omega}{2k_B T} }\right) \left | {\beta _{\uparrow q}(t) }\right|^2 } \;.
\end{equation}

\noindent The integration is straightforward, since the free evolution of the displacements $\beta _{\uparrow q}(t)$ is now known at all times through Eq.({\ref{eq15}), while each rf-pulse contributes a simple change of sign in all displacements. Fig.~\ref{fig1} shows the result for representative model parameters, cycle periods and temperatures in an ohmic environment, $d=1$ [the superohmic $d=3$ case is qualitatively similar]. Should the time reversal of the bath correlation $\eta(t)$ be exact, the maxima would occur precisely at times $t_n=2n\Delta t$, since the derivative of $\Gamma$ is null in the initial state. Because the reversal is only partial, the maxima of $\eta(t)$ are seen to shift gradually, with each cycle, toward the midpoint of the interval between consecutive pulses. Moreover, as the temperature increases, the read-out values at $t_n=2n\Delta t$ tend to become comparable to the lowest values in a cycle. One may infer that shifting the read-out times by $\sim +\Delta t/2$, so as to take advantage of these maxima, may yield a slight improvement of the overall outcome.\\

It turns out, in fact, that a minor rearrangement of the protocol results in a more significant gain. Consider the following version of the idealized bang-bang cycle:  i) propagation under Hamiltonian (\ref{eq1}) for a duration $\Delta t/2$, starting at time $t_n$; ii) first $\pi$-pulse and interchange of qubit eigenstates states at time $(t_n + \Delta t/2)$; iii) evolution under Hamiltonian (\ref{eq1}) from time $(t_n + \Delta t/2)$ to time $t_n +3 \Delta t/2$; iv) second $\pi$-pulse and interchange of qubit eigenstates at time $t_n+3\Delta t/2$; v) readout at $t_{n+1}=t_n+2\Delta t$. This sequence is known in the literature on NMR decoupling methods as a symmetrized Carr-Purcell protocol [see ref.\cite{Viola} for the quantum computation context], and was shown in the framework of the {\it average Hamiltonian theory} [AHT] to yield an improvement in decoupling [suppression of decoherence] of order $\sim {\cal O}(\omega_c \Delta t)^4$ over the standard sequence. Since in the AHT this result is derived directly for the stroboscopic coherence [proportional to our bath correlation factor] at the end of each cycle, we find it instructive to review here a few additional details supplied by the exact coherent-product solution to the associated relaxation problem. 

As for the standard protocol, it can be verified that under this symmetrized sequence the qubit coherence also displays a maximum between any two consecutive rf-pulses. But since the read-out is scheduled now halfway between two pulses, one can expect that the corresponding output values fall close to the coherence maxima following the second pulse in each cycle. In addition, we can anticipate that the symmetrized protocol also benefits from the halved lag [$\Delta t/2$] between the initial configuration to be preserved and the first applied rf-pulse, which induces the first revival. Similarly to an increase in the cycle frequency, this reduced lag leads to faster clipping of the decoherence periods within each cycle, hence again to higher coherence maxima and higher read-out values. 

For a quantitative assessment of these effects, let us resort once more to solution (\ref{eq3}). If the read-out times are $t_n=2n\Delta t$, the bath displacements between consecutive $\pi$-pulses read as follows: 

\noindent i)  for $(t_n-\Delta t/2)< t < (t_n+\Delta t/2)$: $\beta _{\uparrow (\downarrow) q} (t) = \left[ {\beta _{\downarrow (\uparrow) q} (t_n-\Delta t/2-0 ) \pm \chi _q /\hbar \omega _q } \right]\exp [ - i\omega _q (t - t_n+\Delta t/2 )] \mp \chi _q /\hbar \omega _q $; 

\noindent ii) for $(t_n+\Delta t/2) < t < t_n+3\Delta t/2$: $\beta _{\uparrow (\downarrow) q} (t) = \left[ {\beta _{\downarrow (\uparrow) q} (t_n +\Delta t/2-0) \pm \chi _q /\hbar \omega _q } \right]\exp [ - i\omega _q (t - t_n-\Delta t/2 )] \mp \chi _q /\hbar \omega _q $. 

\noindent As before, we assume an unentangled environment in thermal equilibrium, so that $\beta _{\uparrow q} (0)=\beta _{\downarrow q} (0)=0$ and $\beta _{\uparrow q} (t)=-\beta _{\downarrow q}(t)$ for all $t$. Taking this into account leads to a stroboscopic recurrence of the form

\[
\beta _{\uparrow q} (t_{n+1}) = \beta _{\uparrow q} (t_n) e^{ - 2i\omega _q \Delta t}  + \frac{{\chi _q }}{{\hbar \omega _q }}\left({e^{-i\omega_q\Delta t/2}-1}\right)^2 \left( {e^{ - i\omega _q \Delta t}  - 1} \right) \;,
\]

\noindent which solved for $\beta_{\uparrow q}(0)=0$ yields, under a standard density of states, a stroboscopic dissipative factor

\begin{equation}\label{eq39}
\Gamma_{sym} (2n\Delta t)=\lambda \hbar^2 \omega_c \int\limits_{0}^{\infty} { d\omega \left( \frac{\omega}{\omega_c} \right)^{d-1} \exp\left({-\frac{\omega}{\omega_c} }\right) \coth\left({ \frac{\hbar\omega}{2k_B T} }\right) \left | {\beta _{\uparrow q,0}(2n\Delta t)\; \frac{1-\cos(\omega_q \Delta t/2)}{\cos(\omega_q \Delta t/2)} }\right|^2 } \;.
\end{equation} 

\noindent Here $\beta _{\uparrow q,0}(2n\Delta t)$ is again the unperturbed evolved displacement. Note that expression (\ref{eq39}) differs from the standard expression (\ref{eq35}) only through the substitution of the factor $|\tan(\omega_q \Delta t/2)|^2$ by a factor of $|[1-\cos(\omega_q \Delta t/2)]/\cos(\omega_q \Delta t/2)|^2$. Remarkably, the two factors have a qualitatively similar functional dependence on $(\omega_q \Delta t)$, although their periodicity differs from $2\pi$ to $4\pi$, respectively. Quantitatively, on the dominant range $0\leq \omega_q \Delta t \leq \pi$, one has $|\tan(\omega_q \Delta t/2)|^2 \geq |[1-\cos(\omega_q \Delta t/2)]/\cos(\omega_q \Delta t/2)|^2$ and this proves sufficient to render $\Gamma_{strob}(2n\Delta t) \geq \Gamma_{sym}(2n\Delta t)$, hence $\eta_{strob}(2n\Delta t) \leq \eta_{sym}(2n\Delta t)$. The magnitude of this effect has been evaluated by numerical integration of the stroboscopic expression (\ref{eq39}) and the results are displayed in Fig.~\ref{fig2}, alongside the corresponding standard output. In addition, a numerical integration of expression (\ref{eq38b}) for the corresponding exact displacements at intermediate times shows clearly [Fig.~\ref{fig1}] the emergence of the improved suppression of decoherence from the actual revival of coherence induced by each rf-pulse. 

As expected, the qubit coherence is always better preserved under the symmetrized protocol. In addition, the symmetrized outcome is considerably more stable under temperature changes, with coherence variations $\leq 10^{-4}$ compared to $\sim 10^{-2}$ under the standard protocol. The absolute magnitude of the improvement generally does not exceed $5 \%$ [over time] at identical cycle periods and temperatures in an ohmic environment, but may raise above $\sim 10\%$ in the superohmic case [not shown]. However, the true advantage of the symmetrized sequence can be better appreciated in the high precision regime. Indeed, as can be observed from Fig.~\ref{fig2}, in order to maintain the errors in the bath correlation $\eta$ under, e.g., $\sim 0.1\%$ over an extended period of time [say, at least an order of magnitude longer than the typical decoherence time $\sim \omega_c^{-1}$], a standard sequence must be applied at a frequency substantially higher than $\omega_c$. To give a semiquantitative reference, under the choice of model parameters employed in Fig.~\ref{fig2} the rf-frequency must be at least $ 4\omega_c$ at $T=1K$ and at least $6\omega_c$ at $T=100K$. The symmetrized protocol is seen to deliver a similar precision with a sizable reduction in frequency, in an essentially temperature-independent manner. E.g., under the same model parameters the necessary frequency amounts to only $\sim 2\omega_c$ at both $T=1K$ and $T=100K$. This reduction in the operating frequency increases as the precision bounds on the correlation $\eta_{\uparrow\downarrow}$ increase. As a result, a symmetrized protocol applied at $T=100K$ with a frequency of $2.5\;\omega_c$ can perform better [errors $\leq 10^{-4}$] than a standard protocol operating at $T=1K$ with a frequency of $10\omega_c$ [errors $\leq 10^{-3}$].

\section{Conclusion}

In principle, the dynamics of a quantum register on a direct sum of DFSs may remain decoherence free even when the reduced register state is not restricted to a single DFS and involves considerable entanglement with the environment. The present paper provides explicit conditions for the occurrence of this effect in a large class of thermal coherent-product distributions [Eq.(\ref{eq13})], supported by quantum registers with DFSs under interactions linear in the coordinates of a harmonic environment [Eq. (\ref{eq7})]. Such coherent-product states represent distributions on direct sums of register DFS, nontrivially entangled at all times with a statistical superposition of environmental Gaussian states. In particular, any register distribution on a direct sum of DFS develops precisely into a coherent-product distribution when brought in contact with an initially uncorrelated thermal environment. On the other hand, in the zero-temperature, pure state limit these states become self-consistent [but not soliton bearing] Davydov coherent-product states.

Our study was prompted by the straightforward observation that the total state described by Eq.(\ref{eq13}) yields a unitary register evolution in the special case when the associated bath displacements coincide with the stationary displacements characteristic of each register DFS. Here we showed that the set of coherent-product states that propagate the reduced register state in an unperturbed, decoherence-free manner is considerably wider. These states require an appropriate density of degenerate environmental modes such that i) the displaced harmonic environments experienced within contributing DFSs have identical energy reference points [Eqs.(\ref{eq23}a) or (\ref{eq7a6}a)]; ii) the environment Gaussian distributions can be adequately synchronized through their Gaussian displacements [Eqs. (\ref{eq24})]; and iii) the amplitude of the environment mediated correlations between different DFS is finite  [expression (\ref{eq26}) must be finite]. The latter requirement implies that the displaced environment vacua seen by contributing DFS must be unitarily equivalent. Somewhat unexpected, an ohmic environment cannot support such states in single qubits or registers with weak collective decoherence, but is marginally effective for linear registers with individual decoherence. A superohmic environment, on the other hand, is considerably more efficient in sustaining entangled, decoherence-free distributions in a variety of registers, including registers with weak collective decoherence. However, since the amplitude of correlations between distinct DFS compatible with such a propagation decreases significantly with increasing temperature, nontrivial entangled decoherence-free states likely entail a low temperature environment. The simplest examples of joint register-environment states supporting multi-DFS decoherence free register propagation are Davydov ansatz pure states with properly phased coherent environment components. 

The practical worth of these distributions depends to a large extent on one's ability to generate and control a proper phasing of the environmental entanglement. It turns out that qubit flipping through radiofrequency $\pi$-pulses, as in bang-bang suppression of decoherence, provides an elementary procedure for such a manipulation, which amounts to a change of sign of the associated Gaussian displacements. As a corollary, examination of the bang-bang cycle in terms of the coherent-product states (\ref{eq13}) reveals that each rf-pulse in the protocol causes a revival of coherence. It also follows that the well-known time-symmetric version of the cycle takes advantage of these revivals in a natural way, by approximately matching the read-out times to the times of maximum qubit coherence between consecutive pulses. As a result, high precision control becomes possible with cycle frequencies reduced by a factor of about $2 \div 4$ and with a temperature sensitivity diminished by two orders of magnitude. 

Returning to DF coherent-product distributions, let us assume the availability of a realistic preparation procedure. The coherent concatenation of distinct register DFSs effected by such states immediately suggests the possibility of expanding the capacity for noiseless storage beyond the limit set by parallel [incoherent] storage on the same set of DFSs. Consider, as a simple example, a 3-qubit register with weak collective decoherence as described in Sec.VB. The register Hilbert space decomposes into two 1-dimensional DFS, corresponding to each of the states $|\downarrow\downarrow\downarrow\rangle$ and $|\uparrow\uparrow\uparrow\rangle$, and two 3-dimensional DFSs spanned by the basis sets $\{ |\downarrow\downarrow\uparrow\rangle, |\downarrow\uparrow\downarrow\rangle\, |\uparrow\downarrow\downarrow\rangle \}$ and $\{ |\downarrow\uparrow\uparrow\rangle, |\uparrow\downarrow\uparrow\rangle\, |\uparrow\uparrow\downarrow\rangle \}$, respectively. According to the usual theory of DF propagation, each of the 3-dimensional DFSs supports noise-protected storage of one encoded qubit or qutrit. In contrast, a coherent-product state, e.g., a Davydov state of the type described in Sec.IIIB, reading 

\[
|\Psi\rangle = \left[{a_{\downarrow\downarrow\uparrow}|\downarrow\downarrow\uparrow\rangle + a_{\downarrow\uparrow\downarrow} |\downarrow\uparrow\downarrow\rangle\ + a_{\uparrow\downarrow\downarrow} |\uparrow\downarrow\downarrow\rangle }\right] \otimes |\beta_1(t)\rangle + \left[{a_{\downarrow\uparrow\uparrow} |\downarrow\uparrow\uparrow\rangle + a_{\uparrow\downarrow\uparrow} |\uparrow\downarrow\uparrow\rangle\ + a_{\uparrow\uparrow\downarrow} |\uparrow\uparrow\downarrow\rangle }\right] \otimes |\beta_2(t)\rangle 
\]

\[
\equiv a_1 |\Psi_1\rangle \otimes |\beta_1(t)\rangle + a_2 |\Psi_2\rangle \otimes |\beta_2(t)\rangle\;\;,
\]

\noindent allows the noiseless storage of two {\it entangled }encoded qubits or qutrits [that is, of a genuine two-qubit or two-qutrit memory] when the coherent environment states $|\beta_{1,2}(t)\rangle$ have a finite, time-independent overlap, $\langle\beta_1(t)|\beta_2(t)\rangle = const. \neq 0$. Notably, the latter condition makes it possible to extract the relative phases of states belonging to different DFSs [$|\Psi_{1,2}\rangle$ above] solely by means of register measurements, provided the overlap $\langle\beta_1|\beta_2\rangle$ can be determined independently. For instance, one may use a test state of the form 

\[
|\Phi\rangle = \frac{1}{\sqrt 2}\left[{ |\downarrow\downarrow\uparrow\rangle\otimes |\beta_1(t)\rangle +  |\downarrow\uparrow\uparrow\rangle\otimes |\beta_2(t)\rangle }\right]\;\;,
\]

\noindent and projective measurements on the register states $|\phi\rangle = (1/\sqrt 2) [|\downarrow\downarrow\uparrow\rangle + |\downarrow\uparrow\uparrow\rangle ]$ and $|\phi'\rangle = (1/\sqrt 2) [|\downarrow\downarrow\uparrow\rangle + i |\downarrow\uparrow\uparrow\rangle ]$ to extract the desired overlap from the corresponding probabilities $p_{\phi} \equiv \langle \phi| Tr_B [|\Phi\rangle \langle \Phi |] |\phi \rangle = 1/2 + Re\langle\beta_1|\beta_2\rangle$ and $p_{\phi'} \equiv \langle \phi'| Tr_B [|\Phi\rangle \langle \Phi |] |\phi' \rangle = 1/2 + Im \langle\beta_1|\beta_2\rangle$, respectively. Similar register measurements on any encoding state $|\Psi \rangle$ will then allow a full read-out of the information stored in the amplitudes $a_{jkl}$. As usual for a mixed register state, a complete set of such measurements can be obtained from, e.g., projective measurements on each of the states in the computational basis, plus measurements on all entangled superpositions of pairs of basis states of the form $(1/\sqrt 2) [|\alpha\rangle + |\beta\rangle]$ and $(1/\sqrt 2) [|\alpha\rangle + i |\beta\rangle ]$. In general, as shown in Sec.VB, noiseless coherent-product superpositions of DFS under weak collective decoherence enable a doubling in size of the code space, corresponding to the addition of an extra encoded qubit.  Note however that the advantage of coherent-product encoding becomes fully apparent on mixed distributions, where the set of independent storage liberties [matrix elements] becomes considerably larger than in the pure state case.

We conclude by pointing out that the environment-entangled {\it unperturbed} evolution discussed in this paper should be understood as one particular case of environment-entangled {\it unitary} dynamics. For instance, conditions (\ref{eq23}b) or system (\ref{eq24}) suffice to guarantee a unitary, quasi-unperturbed propagation of $\rho _R \left( t \right)$. This is because the stationary phase term $F_0 $ in Eq. (\ref{eq19}) is separable into contributions from individual, orthogonal DFS, and contributes merely {\it stationary} energy shifts to the unperturbed Hamiltonian. Furthermore, it is also possible to obtain sufficient conditions for the unitary propagation of $\rho _R( t)$ by requiring only that the dissipative factors of all $\eta _{J\nu, J'\nu'} $ be stationary [$d\Gamma_{J\nu, J'\nu'}/dt=0$] and that all phase factors be separable, which amounts in fact to $d\Phi _{J\nu, J'\nu'}/dt=0$. The unperturbed Hamiltonian is modified then by {\it time-dependent} energy shifts and the bath displacements must satisfy the constraints

\[
\sum\limits_{\scriptstyle q \atop 
  \scriptstyle \omega _q  = \omega  \ne 0 } {\left( {\beta_{J\nu,q}^0 - \beta_{J'\nu',q}^0} \right) \left( {\mu _{Jq}^*  - \mu_{J'q}^* } \right) }  = 0\;,
\]

\noindent and 

\[
\sum\limits_{\scriptstyle q \atop 
  \scriptstyle \omega _q  = \omega  \ne 0 \hfill} {\left( {\mu_{Jq}^*\beta_{J'\nu',q}^0  - \mu_{J'q}^* \beta _{J\nu,q}^0 } \right) }  = 0\;.
\]

\noindent The requirement that the stationary dissipative factor $\Gamma_{J, J'}^0$ be finite remains, of course, unchanged.\\ \\

\acknowledgments
The author is grateful to L.Viola for pointing out the {\it average Hamiltonian} interpretation of the time-symmetric bang-bang protocol discussed in Sec.VI, as well as its status among NMR decoupling methods.

\begin{center}
\bf Figure captions
\end{center}

\begin{figure}
\caption{Time-dependence of qubit decoherence in an Ohmic environment under standard [hollow symbols] and symmetrized [solid symbols] sequences of rf-pulses, for two typical temperatures corresponding to $k_B T/ \hbar \omega_c = 0.01$ [squares] and $k_B T/ \hbar \omega_c = 1.0$ [circles]. For a cut-off frequency $\omega_c \approx 10^{13}s^{-1}$, the respective temperatures are $T=1K$ and $T=100K$. Time is given in units of the read-out [cycle] period $\tau_{cycle}=2\Delta t$, and the model scaling constant was set to $\lambda \hbar^2 \omega_c=0.25$. }
\label{fig1}
\end{figure}

\begin{figure}
\caption{Errors in qubit decoherence as a function of cycle frequency under standard [Eq. (\ref{eq35}), hollow symbols] and symmetrized [Eq. (\ref{eq39}), solid symbols] sequences applied for a time t, at temperatures $T=1K$ [squares] and $T=100K$ [circles]. Other parameters as in Fig.~\ref{fig1}. Each point represents the read-out after a number of $t/\tau_{cycle}$ cycles. }
\label{fig2}
\end{figure}

\end{document}